\pdfoutput=1

\documentclass[preprint,3p,twocolumn]{elsarticle}

\usepackage{lineno,hyperref}
\modulolinenumbers[5]

\journal{Nuclear Instruments and Methods in Physics Research A}
\newcommand{\gm}{$(g$\,$-$\,$2)$}
\newcommand{\pb}{PbF$_2$}

\begin{document}
\begin{frontmatter}
\title{Studies of an array of \pb\  \v{C}erenkov crystals with large-area SiPM readout}

\author[uw]{A.\,T.~Fienberg}

\author[uw]{L.\,P.~Alonzi\fnref{panote}}
\fntext[panote]{now at University of Virginia,  USA}

\author[frascati,messina]{A.~Anastasi}
\author[cornell]{R.~Bjorkquist}
\author[udine,uudine]{D.~Cauz}
\author[uky]{R.~Fatemi}
\author[frascati,pisa]{ C.~Ferrari}
\author[frascati,pisa]{A.~Fioretti}
\author[cornell]{A.~Frankenthal}
\author[frascati,pisa]{ C.~Gabbanini}
\author[cornell]{L.\,K.~Gibbons}
\author[jmu]{K.~Giovanetti}
\author[uva]{S.\,D.~Goadhouse}
\author[uky]{W.\,P.~Gohn}
\author[uky]{T.\,P.~Gorringe}

\author[uw]{D.\,W.~Hertzog}

\author[napoli,unapoli]{M.~Iacovacci}
\author[uw]{P.~Kammel}

\author[uw]{J.~Kaspar\corref{jknote}}
\cortext[jknote]{Corresponding author: Tel.: +1-206-543-2996.}
\ead{kaspar@uw.edu}

\author[uw]{B.~Kiburg\fnref{bknote}}
\fntext[bknote]{now at Fermi National Accelerator Laboratory, USA}

\author[shangai1,shangai2]{L.~Li}
\author[napoli]{S.~Mastroianni}
\author[udine,uudine]{G.~Pauletta}
\author[uw]{D.\,A.~Peterson}
\author[uva]{D.~Po\v{c}ani\'{c}}
\author[uw]{M.\,W.~Smith}
\author[cornell]{D.\,A.~Sweigart}

\author[uky]{V.~Tishchenko\fnref{vtnote}}
\fntext[vtnote]{now at Brookhaven National Laboratory, USA}

\author[frascati]{G.~Venanzoni}
\author[uw]{T.\,D.~Van\,Wechel}
\author[uw]{K.\,B.~Wall}

\author[uw]{P.~Winter\fnref{pwnote}}
\fntext[pwnote]{now at Argonne National Laboratory, USA}

\author[uw,osaka]{K.~Yai}

\address[uw]{University of Washington, Box 351560, Seattle, WA 98195, USA}
\address[cornell]{Cornell University, Ithaca, NY 14850, USA}
\address[frascati]{Laboratori Nazionali Frascati dell' INFN, Frascati, Italy}
\address[udine]{INFN, Sezione di Trieste e G.C. di Udine, Udine, Italy}
\address[napoli]{INFN, Sezione di Napoli, Napoli, Italy}
\address[pisa]{Istituto Nazionale di Ottica del C.N.R., UOS Pisa, Pisa, Italy }
\address[messina]{Dipartimento di Fisica e di Scienze della Terra dell'Universit\`a di Messina, Messina, Italy}
\address[unapoli]{Universit\`a di Napoli, Napoli, Italy}
\address[uudine]{Universit\`a di Udine, Udine, Italy}
\address[uky]{University of Kentucky, Lexington, KY 40506, USA}
\address[uva]{University of Virginia, Charlottesville, VA 22904, USA}
\address[jmu]{James Madison University, Harrisonburg, VA 22807, USA}
\address[shangai1]{Shanghai Jiao Tong University, Shanghai, China}
\address[shangai2]{Shanghai Key Laboratory for Particle Physics and Cosmology, Shanghai, China}
\address[osaka]{Osaka University Graduate School of Science, Osaka, Japan}

\begin{abstract}
The electromagnetic calorimeter for the new muon \gm~experiment at Fermilab will consist of arrays of \pb\ \v{C}erenkov crystals read out by large-area silicon photo-multiplier (SiPM) sensors.  We report here on measurements and simulations using 2.0\,--\,4.5\,GeV electrons with a 28-element prototype array.  All data were obtained using fast waveform digitizers to accurately capture signal pulse shapes versus energy, impact position, angle, and crystal wrapping. The SiPMs were gain matched using a laser-based calibration system, which also provided a stabilization procedure that allowed gain correction to a level of $10^{-4}$ per hour. After accounting for longitudinal fluctuation losses, those crystals wrapped in a white, diffusive wrapping exhibited an energy resolution $\sigma/E$ of  $(3.4\pm0.1)\,\%/\sqrt{E/\mathrm{GeV}}$, while those wrapped in a black, absorptive wrapping had $(4.6\pm0.3)\,\%/\sqrt{E/\mathrm{GeV}}$. The white-wrapped crystals---having nearly twice the total light collection---display a generally wider and impact-position-dependent pulse shape owing to the dynamics of the light propagation, in comparison to the black-wrapped crystals, which have a narrower pulse shape that is insensitive to impact position.
\end{abstract}

\end{frontmatter}

\section{Introduction}

The new muon \gm~experiment~\cite{Carey:2009zzb} at Fermilab will require 24 electromagnetic calorimeter stations placed on the inside radius of a magnetic storage ring. The muon precession frequency data is obtained from the decay of 3.1\,GeV/$c$ muons repeating many $\sim\!700\,\mu$s ``fills.''  The detectors must accurately measure the hit times and energies of the positrons, which curl to the inside of the ring following muon decay.  For maximum acceptance, the calorimeters are located partly within the storage ring's highly uniform 1.45\,T magnetic field and extend inward radially to a region where the field falls to $\approx0.8$\,T.  A rigorous material selection and evaluation process is required for candidate absorber and readout components to avoid perturbation to the uniformity of the field.

Demands on the calorimeter and readout design are based on the unusual nature of the \gm~measurement, where gain instabilities and pulse pileup introduce major systematic
uncertainties. Instantaneous rates are expected to exceed a few MHz per station at the beginning of any storage ring fill and drop, on average, exponentially by more than 4 orders of magnitude over the measuring period.  Over this measuring period, temporal gain stability of better than 0.1\,\% must be maintained.  At all times, two-pulse resolution for events separated by 5\,ns or more in the same detectors is required.  These demands have led to a limited number of design concepts~\cite{Sedykh:2000ex,McNabb:2009dz}. Here, we describe a detailed evaluation of a new option, which has been selected for the \gm~experiment.

A calorimeter station will consist of 54 lead fluoride (\pb) crystals in a 6 high by 9 wide array, with each crystal read out on the rear face using a large-area SiPM coupled directly to the crystal surface.  While \pb\  calorimeters have not been extensively used in the past%
\footnote{The A4 Collaboration published a number of important technical papers; see \cite{Achenbach:2001kf,Achenbach:1998yv}},
their properties are particularly well suited to the needs of the \gm~experiment. \pb~has very high density (7.77\,g/cm$^3$), a 9.3-mm radiation length, and a Moli\`ere radius of $R_M^E = 22$\,mm for energy deposition. An equivalent to Moli\`ere radius for \v{C}erenkov photons is a cylinder inside which 90\,\% of photons are generated. The radius was measured as $R_M^C = 18$\,mm~\cite{Anderson:1989uj}. The fast nature of the purely \v{C}erenkov radiation aids in reducing  pileup. In fact, the intrinsic pulse width from photon arrivals is affected noticeably by the choice of wrapping, as we will detail below.

The relatively new development of SiPMs as light transducers~\cite{Renker:2006ay} has considerable advantages compared to PMTs, albeit with several distinct challenges.  Their compact nature provides freedom in the mechanical design of the calorimeter housing so they can be mounted in tight geometries.  They operate in high magnetic fields without degradation and, important for \gm, they do not perturb the magnetic field as long as a suitable choice of the electronics support components is made.  The well-known challenges include the need for temperature and bias control stability during operation.

In principle, the energy resolution of a \pb\,--\,SiPM combination should be several times better than the lead-scintillating fiber (SciFi) sampling calorimeter used in the previous \gm\ experiment, as long as the light can be efficiently detected by the SiPMs.  In SciFi-based calorimeters, light can be concentrated by suitable tapered guides and directed to a smaller area light sensor with minimal loss~\cite{Simon:1993ub}.  This feature does not exist for crystals where the light is well mixed and uniformly fills the downstream face at large incindent angles. The challenge is to enable the light, which propagates dominantly via total internal reflection, to efficiently escape from \pb, with its high refractive index of 1.8, to the SiPM sensitive surface.  Even in their larger commercial formats, a single SiPM is small compared to typical calorimeter crystal dimensions. In the device described below, the SiPM employed covers only 23\,\% of the 25\,$\times$\,25\,mm$^2$ crystal face.

\begin{figure}[t]
    \includegraphics[width=1.0\columnwidth]{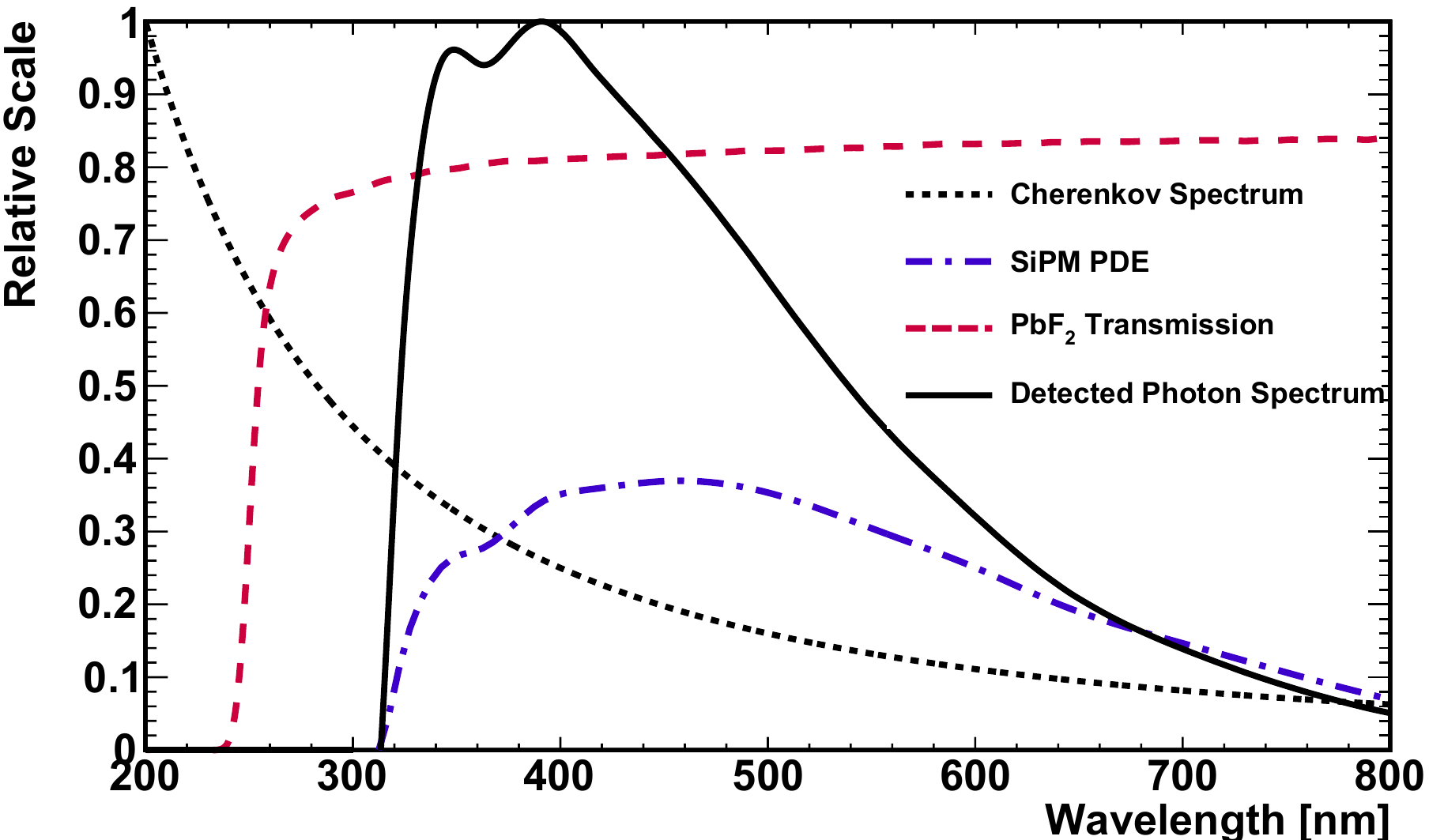}

    \caption{For illustration, the three inputs that determine the relative detected photon spectrum (black full) vs. wavelength are shown: \v{C}erenkov spectrum (black dotted), \pb\ nominal transverse optical transmission across a 25-mm crystal measurement (red dashed), and SiPM quoted PDE at a typical operating over-voltage (blue dash-dot).}
    \label{fg:lightyield}
\end{figure}

Fig.~\ref{fg:lightyield} displays the typical \v{C}erenkov light spectrum behavior, the crystal transmission, and the SiPM photo-detection-efficiency (PDE), as a function of wavelength.  The product of the three gives the relative detected photon spectrum, which aids in guiding the design of the optical system.

\begin{figure*}[t]
    \includegraphics[width=\textwidth]{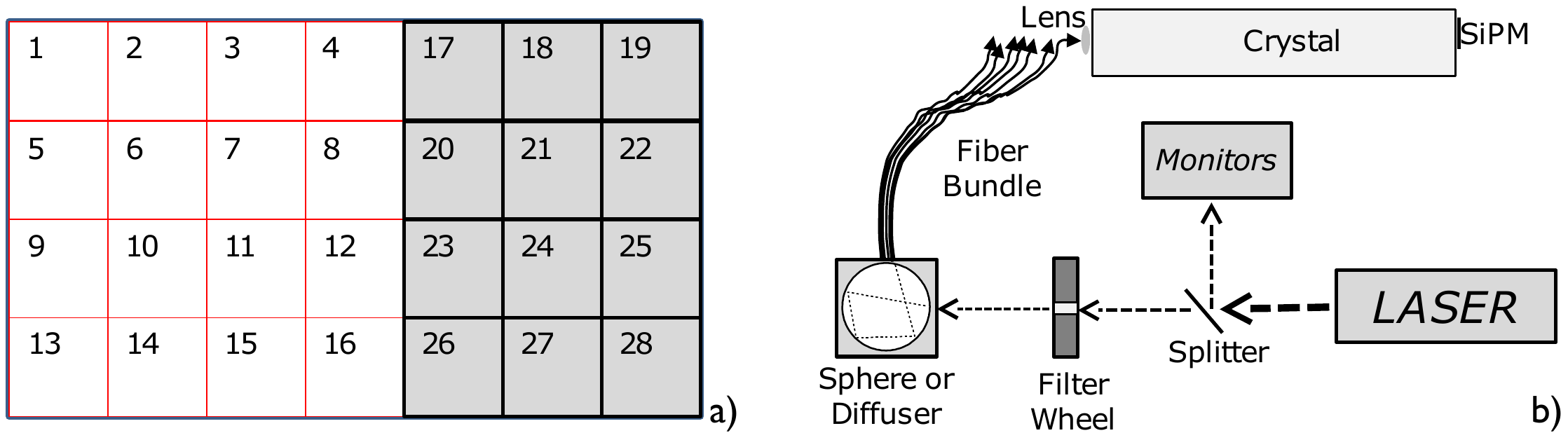}

    \caption{\emph{(a)} Front view of the array of 28 \pb\  crystals.  Elements 1\,--\,16 are wrapped in white Millipore paper, while elements 17\,--\,28 are wrapped in black Tedlar. \emph{(b)} Schematic diagram of the laser calibration system used.  Light from each fiber in the distributed bundle is directed through a lens into the front face of a single crystal.}
    \label{fg:Setup}
\end{figure*}

\section{Setup for Electron Beam Studies}

To characterize the performance and properties of this detector, we completed a study of a prototype array at SLAC's End Station Test Beam Facility.  The facility provides a well-collimated beam of electrons at a user-defined rate with a typical rate of 5\,--\,10\,s$^{-1}$.  In each precisely timed beam pulse, a Poisson distribution of electrons arrives.  When well tuned, the single-electron beam pulses will be most likely, with the probability of $\sim\!37$\,\%.  The beam radial extent when exiting the last vacuum pipe is expected to be $\sim\!$1\,--\,2\,mm and its position is stable, thus avoiding the need for external wire chambers or start counters.  Energies from 2 to 4.5\,GeV were used in the present study. At each setting the beam energy was known to about 50\,MeV and stable to better than 1\,\%.

The calorimeter prototype tested at SLAC was a 4\,$\times$\,7 array of 2.5\,$\times$\,2.5\,$\times$\,14\,cm$^3$ ($15 X_0$) high-quality \pb~crystals, grown by SICCAS%
\footnote{Shangai SICCAS High Technology Corporation, 1295 Dingxi Rd., Shanghai 200050, China}.
Each crystal in the first four consecutive columns was wrapped in a single, non-overlapping layer of reflective white Millipore\textsuperscript\textregistered\ paper, whereas each crystal in the remaining three columns were wrapped in matte black absorbing Tedlar\texttrademark.
The Millipore Immobilone-P is a polyvinylidene fluoride membrane with 0.45\,$\mu$m pores, and it is a Lambertian (diffusive) mirror. The upstream face for all crystals was left unwrapped to permit the injection of light from a calibration system.
A schematic diagram and numbering scheme used in this paper is shown in Fig.~\ref{fg:Setup}, (a).

\begin{figure}[t]
    \includegraphics[width=1.0\columnwidth]{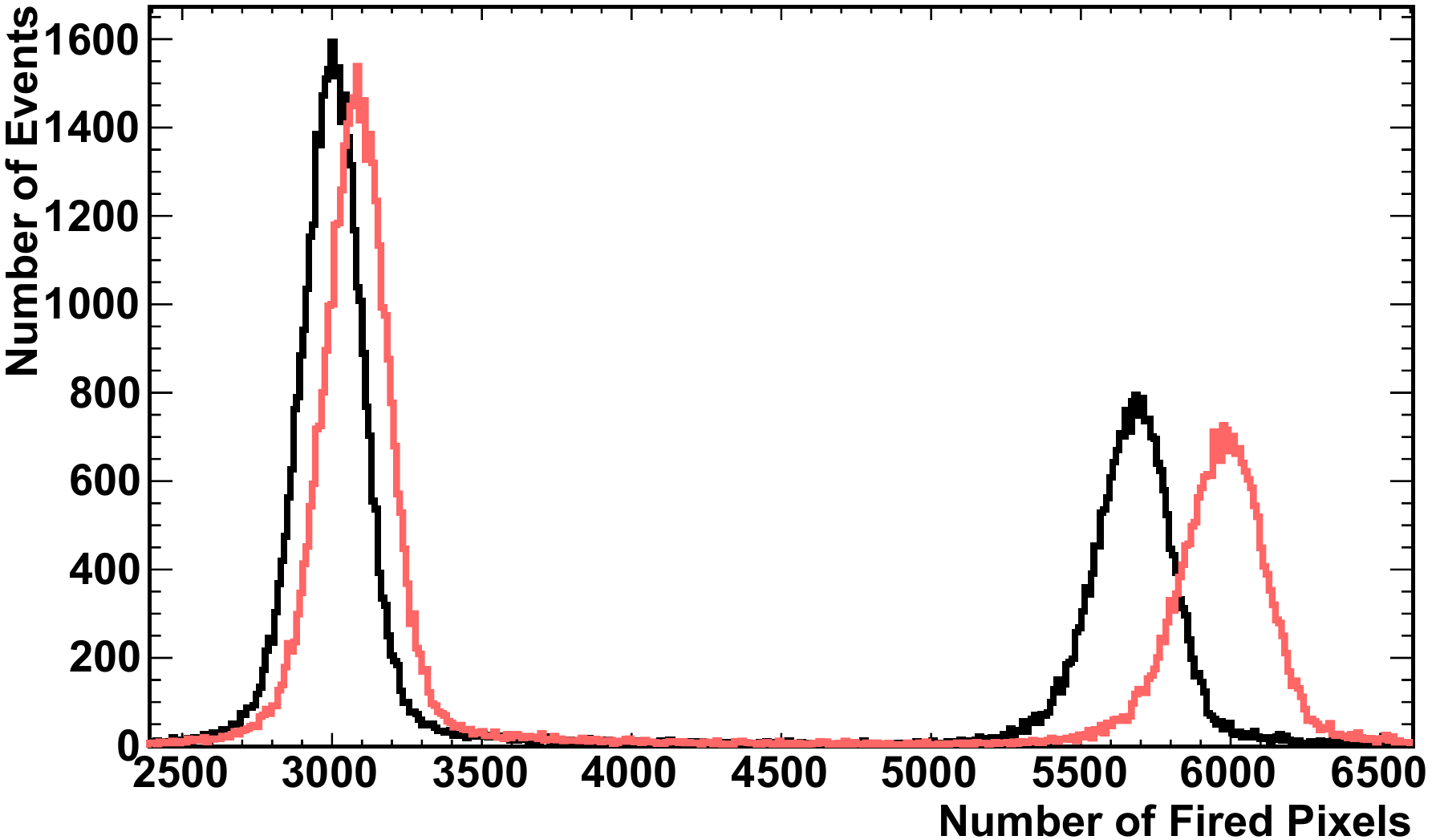}

    \caption{A histogram depicting the response of a single crystal to one or two 3-GeV electrons within a beam bunch.  The raw trace (black) is corrected (light red) to account for pixel saturation.  The correction improves the energy linearity between the first and second peak.}
    \label{fg:CORRECTED_SPECTRUM}
\end{figure}

Each crystal was viewed by a monolithic 16-channel Hamamatsu MPPC%
\footnote{Multi-Pixel Photon Counter Model number S12642-4040PA-50~\cite{Hamamatsu:2014MPPC}.}
(SiPM). The SiPM used has 57,600 50-$\mu$m-pitch pixels in a 1.2\,$\times$\,1.2\,cm$^2$ area, an entrance window made from epoxy resin with the refractive index of 1.55, and was optically matched to \pb~via NuSil LS-5257 optical grease. When a photon strikes a SiPM pixel, it can cause an avalanche that is summed together with the other struck pixels in a linear fashion to produce the overall response. Quenching resistors are intrinsic to the device to arrest the avalanche and allow a fired pixel to recover with a time constant typically in the 10's of ns. The pixel recovery time is very much dependent on the SiPM fabrication properties.  For good near-linear operation, the concept is to have a pixel count that greatly exceeds the highest photon count that would strike the device.  A deviation from linearity at high light levels is caused by pixel saturation, that is, the suppressed ability for a single pixel to have more than one avalanche within a single recovery period. For our crystals, we anticipated approximately 1\,pe/MeV, where pe  (short for photo-electron) represents a converted photon.  The \gm~highest single electron energy is $\sim\!3100$\,MeV,
which implies a maximum pixel occupancy fraction near 5\,\%. Despite the low occupancy fraction, optimal linearity is achieved by applying a correction for pixel saturation, following

\begin{equation}
  \label{eqn:SATURATION}
  N_\mathrm{fired} = N_\mathrm{tot} \left[1-\exp\left(-N_\mathrm{primary}/N_\mathrm{tot}\right)\right] \ .
\end{equation}
Here, $N_\mathrm{fired}$ is the number of pixels that actually fired, $N_\mathrm{tot}$ is the total number pixels a SiPM has, and $N_\mathrm{primary}$ is the number of pixels that would fire if the number of pixels were infinite (and the pixel area vanishing). The formula assumes a uniform illumination which is guaranteed by the optical properties of the \pb~crystal boundaries.
Fig.~\ref{fg:CORRECTED_SPECTRUM} shows a typical energy distribution in one of our crystals from a well-tuned electron beam, which shows peaks for one and two 3-GeV electrons per bunch.  The black trace is the raw response, while the red trace demonstrates the applied pixel saturation correction. The selected SiPM counts multiple thousands of photon hits in a perfectly linear way after the simple analytical correction is applied to compensate for individual pixel saturation.

\begin{figure}[t]
    \includegraphics[width=\columnwidth]{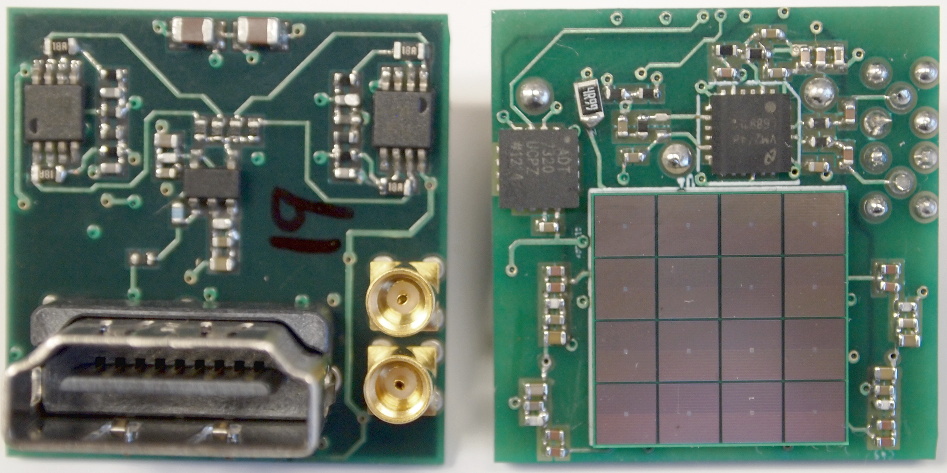}

    \caption{A baseline prototype of a surface-mount 16-channel SiPM soldered on the amplifier board.
    The two MMCX connectors represent the AC coupled differential voltage signal out.
    The common bias voltage in, the board low voltage, and SPI lines to regulate gain
    are supplied through the HDMI connector.}
    \label{fg:SIPM_SURFACE}
\end{figure}

The amplifier board, shown in Fig.~\ref{fg:SIPM_SURFACE}, used to sum up the 16 individual channels is based on a concept of a multi-staged transimpedance op-amp. In the first step, current pulses from 4 SiPM channels are added together and converted into voltage pulses in a fixed-gain transimpedance amplifier. The primary stage is designed around a THS32302 operational amplifier operated in current mode at the constant gain of 600\,$\Omega$, which means that the charge of 1\,pC entering the op-amp within 1\,ns generates a voltage pulse with the amplitude of 0.6\,V. In the second stage, the four partial sums are added together using a THS3201 op-amp operated at unity gain in voltage mode.
The output stage drives an AC coupled differential pair of coaxial cables, and is designed using a LMH6881 digitally controlled variable gain amplifier. The multi-staged design provides a PMT-like pulse width, very high rate tolerance, and excellent gain stability.

\begin{figure}[t]
   \includegraphics[width=1.0\columnwidth]{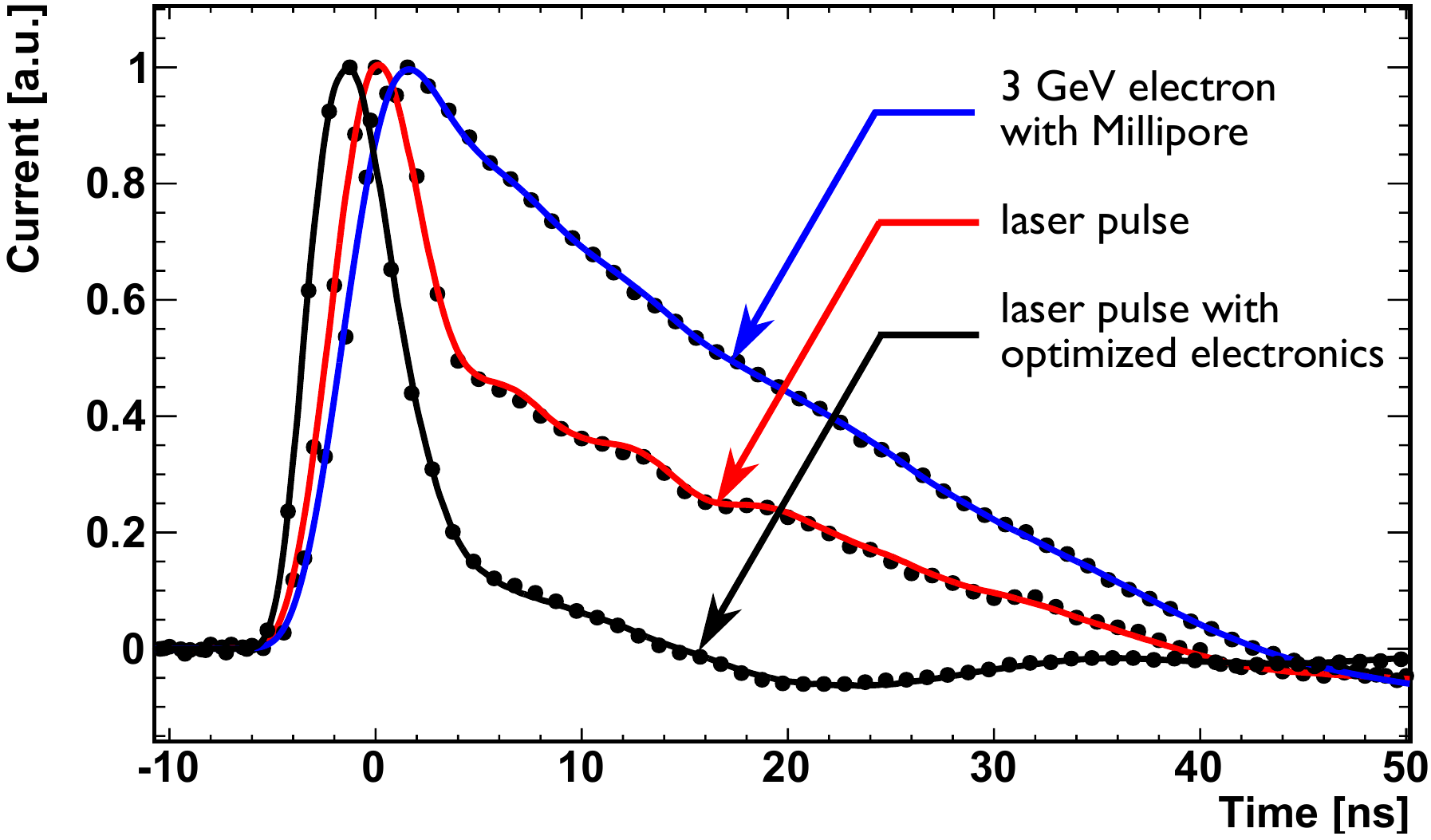}

   \caption{Sample SiPM traces from a 3-GeV electron in a Millipore-wrapped crystal and a laser pulse were recorded by a 1\,GSa/s digitizer. Also shown is a trace from a laser pulse read out by our newest generation of electronics boards. No noise reducing or smoothing filters were applied to these traces.}
   \label{fg:UnfilteredTraces}
\end{figure}

Critical to the studies reported here is the readout of all pulses using high-speed waveform digitizers.
For most studies, the SiPMs coupled to the 16 white-wrapped crystals were digitized using a 12-bit CAEN DT5742 switched capacitor desktop digitizer sampling at 1\,GSa/s, while 8 of the black-wrapped crystals were digitized using a 12-bit Struck SIS3350 digitizer sampling at 0.5\,GSa/s. The digitizers have single ended 50\,$\Omega$ inputs. The differential signal from the SiPM amplifier board was turned into the single-ended one via a balun transformer by Micro Circuits. Fig.~\ref{fg:UnfilteredTraces} depicts sample traces from the CAEN digitizer. The digitized pulse shape allows reliable reconstruction of hit energy, time and position because the shapes of an energy shower and the digitized pulse are directly related.

A high-performance calibration system was used to set the gains of the individual crystal/SiPM elements as well as to monitor the gain stability of each detector throughout the test beam period. Several prototype elements of the overall optical system being designed for the \gm~experiment were tested, including a suite of out-of-beam pin diode monitors, which measured the laser shot-to-shot intensity fluctuations.  The studies reported here relied on a subset of these tools including the light source, a system to precisely control its intensity by attenuation, and a  distribution system to direct light separately to each of the 28 detector elements (See Fig.~\ref{fg:Setup}, (b)).

A PicoQuant LDH-P-C-405M pulsed diode laser drove the calibration system.  It featured a pulse width $<$600\,ps at a wavelength of 405\,$\pm$\,10\,nm.
The power stability over 12 hours in an environment with temperature stability $\Delta$T (ambient) $<$3\,K is quoted as 1\,\% RMS and  3\,\% peak-to-peak, a performance we verified in the field.  The maximum energy per pulse is $\sim\!500$\,pJ.
The pulsed laser was assembled on an optical bench; the light beam passed in air through various fixed-ratio optical splitters, then through a 6-position neutral-density filter wheel, which could be rotated remotely.  The transmission fractions used were: 100\,\%, 82\,\%, 65\,\%, 45\,\%, 30\,\%, and 20\,\%. The multiple laser intensities allow us to discriminate against effects that scale non-linearly with the number of photons.

The emerging beam next entered either of two tested light diffuser systems: a 2-inch integrating sphere, or a diffuser. The integrating sphere (Thorlabs, mod.~IS200-4, 4 ports) provides a high degree of output uniformity at the price of a higher attenuation~\cite{Anastasi:2014xx}. We used a 1-inch diameter, 20$^\circ$ Circle Pattern Diffuser%
\footnote{Engineered Diffuser\texttrademark\ by RPC Photonics, Rochester, NY, purchased from Thorlabs as model ED1-C20},
preceded by a beam expander consisting of two lenses having 10-mm and 50-mm focal lengths, respectively. The diffuser provided more than 10 times larger transmission efficiency, at the price of a somewhat lower spatial uniformity, which might potentially manifest as non-linearity in our calibration data. Two independent light distribution systems provided the desired redundancy in laser calibration.

A custom fiber bundle transmitted light from either distribution system to the calorimeter front panel. The bundle has thirty optical silica fibers with 0.6-mm diameters each, numerical aperture NA${} = 0.39$, and was produced by VINCI%
\footnote{VINCI fine instruments, Via Ciceruacchio 7, Monterotondo, I-00015 Rome, Italy}.
On the calorimeter front panel, each fiber was secured using SMA connectors to a port in front of each crystal (they were offset slightly from the geometrical center). Aspherical lenses (Thorlabs, mod.~CAY046, $f = 4.6$\,mm) were positioned between the SMA interface and the crystal front face to collimate the light output from the fibers. The setup maximized optical transmission from the calibration system into the \pb~crystals.

\section{Operational Procedures}
\subsection{Pulse Fitting}

The pulse-fitter extracts the pulse-integral (pulse area) and hit time from a digitized trace. The pulse-integral is an effective measure of the number of pixels fired. The pulse-fitter used in this study was based on custom pulse templates for each individual SiPM to be robust against small fluctuations in pulse shape.

Data-driven templates (also known in literature as ``system functions'') were built by averaging more than 10,000 digitizer traces and interpolating between digitized samples within a trace using a cubic spline. Templates $T(t')$ were normalized such that $\int T(t')\,dt' = 1$, and aligned in the time domain so that $t'=0$ corresponds to the pulse maximum, which was interpolated by a parabolic curve across 3 samples---the peak sample and its two neighbors. The function used for fitting traces was of the following form:
\begin{equation}
  f(t) = p_0\cdot T\left(\frac{t-p_1}{p_3}\right) + p_2 \quad .
\end{equation}
The four free parameters of this fit are an overall scale factor ($p_0$), the peak time ($p_1$), the DC offset ($p_2$), and a stretch parameter ($p_3$). The stretch parameter allows the fitter to successfully accommodate small variations in pulse shape that may occur during the experiment, and its value range spans from 0.9 to 1.4. Finally, the pulse-integral is extracted as $p_0 \cdot p_3$.

\begin{figure}[t]
   \includegraphics[width=1.0\columnwidth]{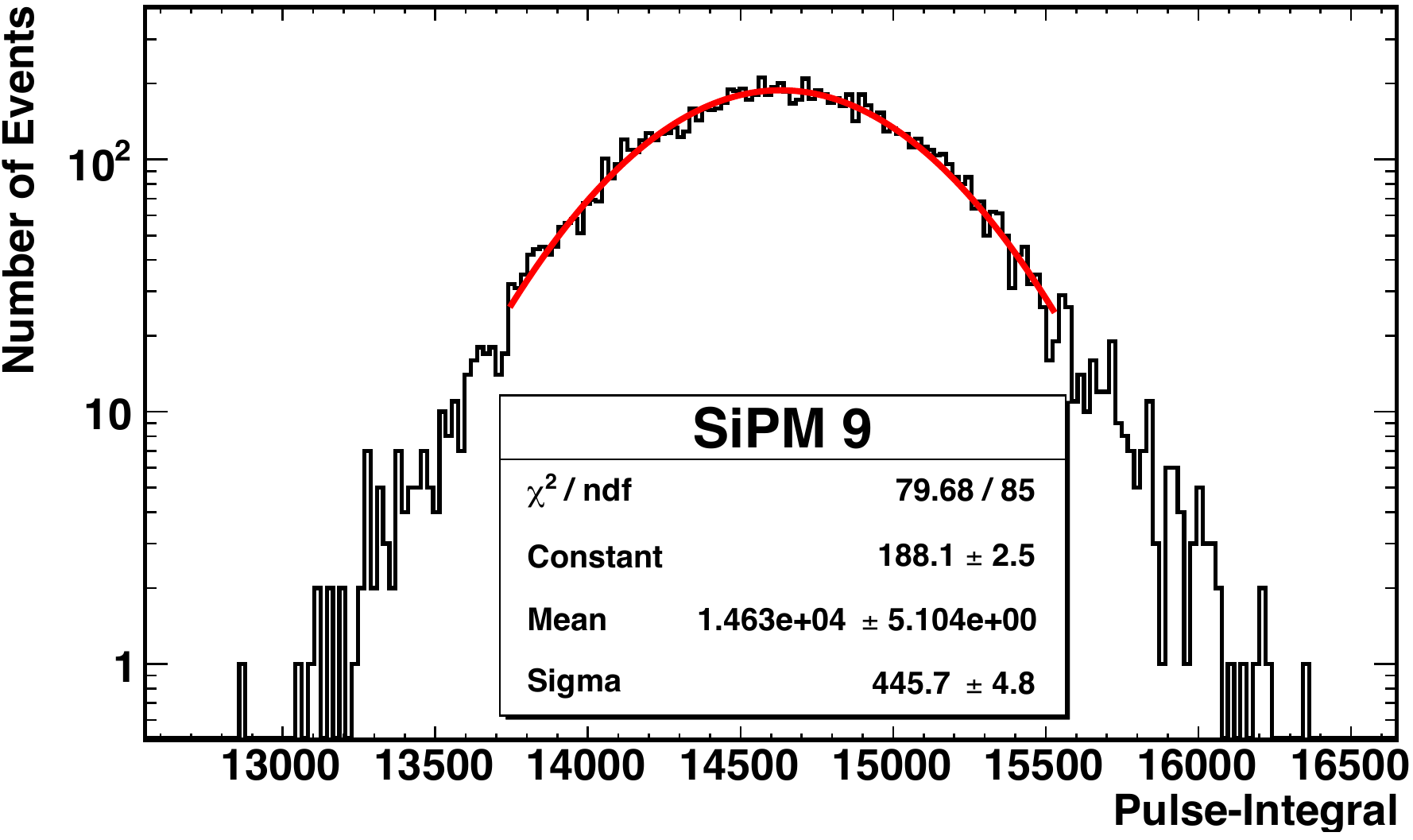}

   \caption{A histogram of pulse-integrals from a single SiPM during a typical laser calibration run with the filter wheel at its open setting (100\,\% transmission). The histogram is fit with a Gaussian to extract the mean and standard deviation.}
   \label{fg:LASER_CALIBRATION_HISTOGRAM}
\end{figure}

To increase numerical stability and eliminate high-frequency noise, all traces were filtered with 10\,ns moving averages before being subjected to the fitter. Since uncertainties of the fit parameters were not used further in the analysis, the artificial correlations possibly introduced or altered by the smoothing procedure were of no concern. The software solution was a work-around for a missing anti-aliasing filter in our setup.

When fitting, a $\chi^2$ minimization was used because it exhibits numerically robust behavior, sufficient region of convergence, and although the pulse-integrals are biased towards lower values, the bias is negligible compared to statistical fluctuations. However, the $\chi^2$ minimization requires defining the statistical uncertainty of the digitized samples.  The best performance---in terms of numerical stability, and number of iterations required to reach the minimum value---was obtained when each digitizer sample was given equal weight in the fit; i.e., all the experimental uncertainties were set to unity. 
The pulse times extracted from the waveform fits were reproducible to better than 50\,ps. A more detailed analysis of timing resolution will be subject of a dedicated paper.

\subsection{Laser Calibration}

The effective number of photo-electrons serves as a proxy for energy deposited into a crystal. The goal of the laser calibration is to convert the pulse-integral from the pulse fit into the number of pixels fired.  After the number of pixels fired is corrected for pixel saturation, it gives the effective number of photo-electrons. The calibration procedure relies on statistical properties of a histogram of many thousands of laser hits collected under stable conditions. The conversion factor from pulse integrals to photo-electrons is the SiPM gain for our purposes in units of pulse-integral/pe.

For a given setting of the neutral-density filter wheel, the pulse-integrals for each SiPM will be normally distributed. An example distribution is shown in Fig.~\ref{fg:LASER_CALIBRATION_HISTOGRAM}. The relationship between the mean and the variance of the normal distribution is determined by the number of incident photons and properties of the individual SiPM.  These properties include the gain, which is the charge delivered by a SiPM pixel when a pixel fires, and the photon detection efficiency (PDE), which is a product of quantum efficiency, the
probability that the charge carrier triggers an avalanche discharge and
the geometrical filling factor. Apart from the geometrical factor, these factors depend on the applied bias voltage.  Finally, the width in Fig.~\ref{fg:LASER_CALIBRATION_HISTOGRAM} can also be affected by external instabilities such as the laser intensity fluctuations during the measuring period.

\begin{figure}[t]
   \includegraphics[width=1.0\columnwidth]{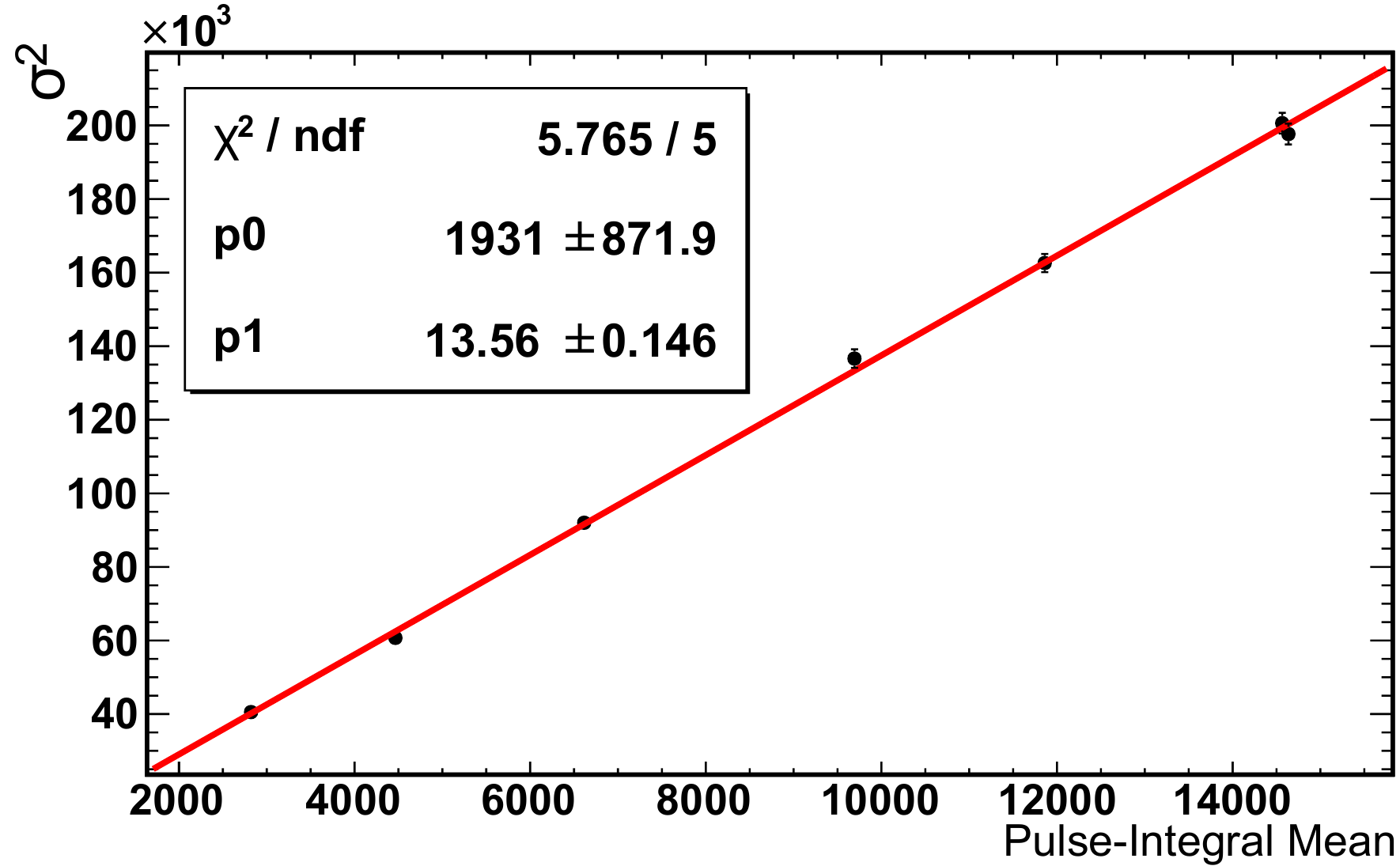}

   \caption{A plot of pulse-integral variance vs. mean as a neutral-density filter wheel is stepped through its positions. The linear term, p1, is equal to the pulse-integral/pe.
     The open filter setting (100\,\% transmission) for this particular SiPM corresponds to approximately 1000 pe.}
   \label{fg:LASER_CALIBRATION_GAIN}
\end{figure}

For a stable laser, the relationship is
\begin{equation}
  \sigma^2 = n_e^2+g\mu \quad ,\quad  (\mu/g \approx N_\mathrm{pe}) ,
  \label{eq:gain-model-linear}
\end{equation}
where $\mu$ is the mean of the pulse-integral distribution, $n_e$ is the amount of electrical noise in the system, and $g$ is the parameter of interest%
\footnote{Undesirable SiPM behaviors such as after-pulsing and cross-talk can further increase the width of the pulse-integral distribution. This causes a systematic underestimate of the number of pe but it has no effect on the linearity of the device.},
pulse-integral/pe.
$N_\mathrm{pe}$ stands for average number of photons registered by a SiPM.
In the case of a more widely varying light source, the relationship can be modified to
\begin{equation}
  \sigma^2 = n_e^2+g\mu+\sigma^2_l\mu^2
  \label{eq:gain-model-quadratic}
  \end{equation}
where $\sigma^2_l$ is the additional relative variance of the light source beyond Poisson statistics%
\footnote{If the distribution of photons incident at the photo-detector from shot-to-shot were normally distributed with mean $m$ and width $\sigma$, the relative variance beyond Poisson statistics is equal to $\sigma^2/m^2 - 1/m$.}.
Thus, by stepping through the filter wheel and extracting the linear term in the $\sigma^2$ vs mean curve, the desired gain constants are determined (Fig.~\ref{fg:LASER_CALIBRATION_GAIN}). Both linear and quadratic fits were performed. When the quadratic term is found statistically significant, the linear term from the quadratic fit (Eq.~\ref{eq:gain-model-quadratic}) is used as the gain value. If the quadratic term is not statistically significant, the gain value used in the analysis comes from the linear fit, i.e., the model described in Eq.~\ref{eq:gain-model-linear}. The switching doesn't pull the fit values, which was verified, but it improves uncertainties. Since the gain and the PDE of a SiPM vary with changes in temperature and bias voltage, the calibration constants had to be tracked through time and continually updated.

\begin{figure}[t]
   \includegraphics[width=1.0\columnwidth]{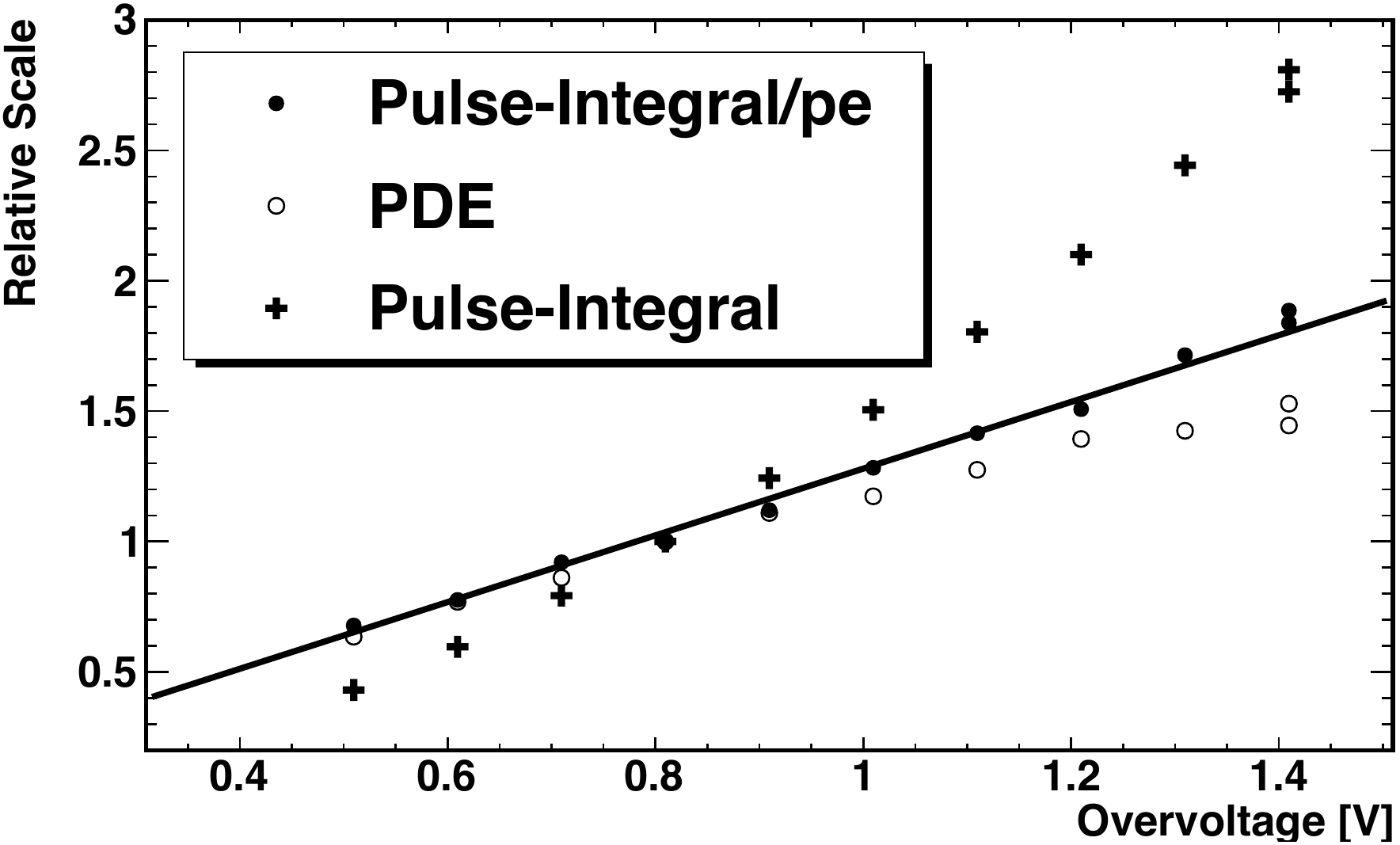}

   \caption{Pulse-integral/pe, PDE, and pulse-integral as a function of over-voltage for SiPM \#11.
     All are normalized to 1 at the run operating voltage of 66.5\,V, which corresponds to an over-voltage of approximately 0.8\,V for this particular SiPM. While the pulse-integral/pe (gain) varies linearly with the over-voltage, the pulse-integral itself increases more rapidly as the product of PDE and gain. }
   \label{fg:BiasVsGain}
\end{figure}

\subsection{Operating Bias Voltage}

SiPMs are operated in the Geiger mode, i.e., the applied bias voltage is greater than the breakdown voltage. The difference between the bias voltage and the breakdown voltage is called over-voltage and controls the SiPM gain, PDE, dark-current, and after-pulsing. The gain scales linearly with over-voltage. The PDE depends on over-voltage, because it drives the probability that a charge carrier triggers an avalanche discharge. The probability is linear for low over-voltages and saturates for higher over-voltages. Both the dark-current and the after-pulsing increase as over-voltage is raised.

While the bias voltage is easy to control, the breakdown voltage, which depends on temperature, cannot be measured directly. It can be inferred indirectly in the following way: the bias voltage is scanned over a range of values,  and a laser calibration is conducted at each step (see Fig.~\ref{fg:BiasVsGain}). The laser calibration procedure disentangles gain and PDE because the pulse-integral is a product of gain and PDE while the effective number of photo-electrons is a measure of relative PDE only. The gain curve (pulse-integral/pe) is fit with a linear model, and the intercept with the bias voltage axis is used as the effective value of breakdown voltage. This effective definition of the breakdown voltage is practical to work with in the field.

\begin{figure}[t]
   \includegraphics[width=1.0\columnwidth]{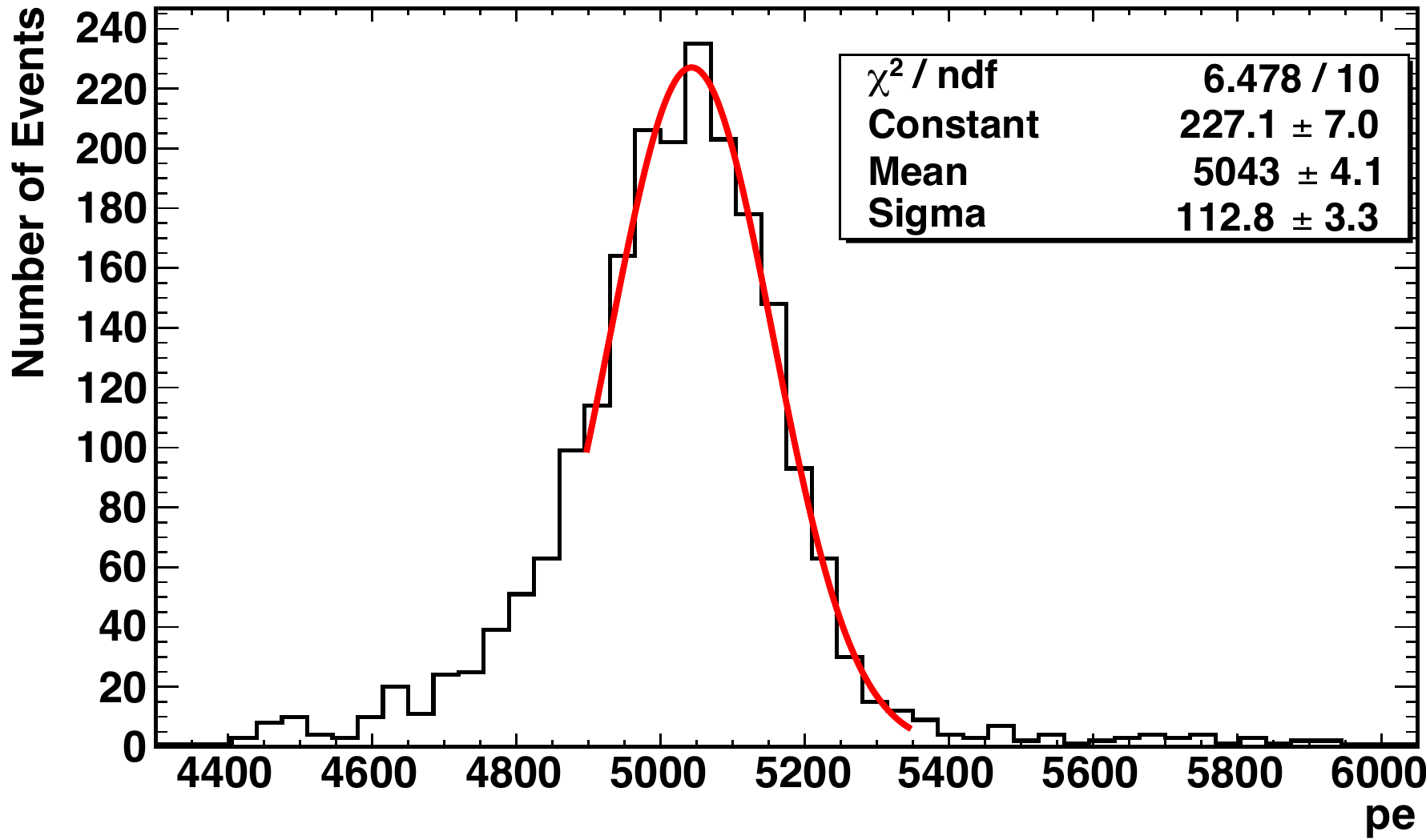}

   \caption{The total number of pe detected among nine white-wrapped crystals arranged in a 3$\times$3 cluster centered on SiPM \#11 when a 3.5-GeV electron impacts the center.}
   \label{fg:SAMPLE_FIT}
\end{figure}

The bias voltages were scanned from 65.8\,V to 67.2\,V. The obtained values of breakdown voltages were around 65.7\,V. The higher than usual value of the breakdown voltages were caused by the warm environment and insufficient cooling. There are multiple thousands of photons in each beam hit or laser shot, which allows lower-than-recommended%
\footnote{The over-voltage value recommended by the manufacturer is the value corresponding to the gain of $1.25 \times 10^6$ and the dark rate for a single cell of about 0.5\,MHz.}
over-voltage values to be used. An operating voltage was set to 66.5\,V, which corresponds to about 0.8\,V over-voltage.  The chosen bias voltage value optimized PDE and gain for the dynamic range of our electronics, allowed us to reconstruct several 3-GeV electrons per beam hit, and reduced the dark-rate.

Within a SiPM, the 16 individual channels share a common bias voltage and charge buffer capacitor through the preamplifier board. Differences in breakdown voltages between pairs of channels are typically less then 50\,mV and the maximum difference between any pair of channels in any of the SiPMs used is 110\,mV. These differences are not important to the studies described here because the number of incident photons is typically in the thousands, effectively averaging the gain of each SiPM over its constituent channels with negligible variation from shot-to-shot. Additionally, the optical properties of the crystals yield a uniform illumination of the SiPMs, ensuring the gain average remains independent of electron impact position.   

\begin{figure}[t]
   \includegraphics[width=1.0\columnwidth]{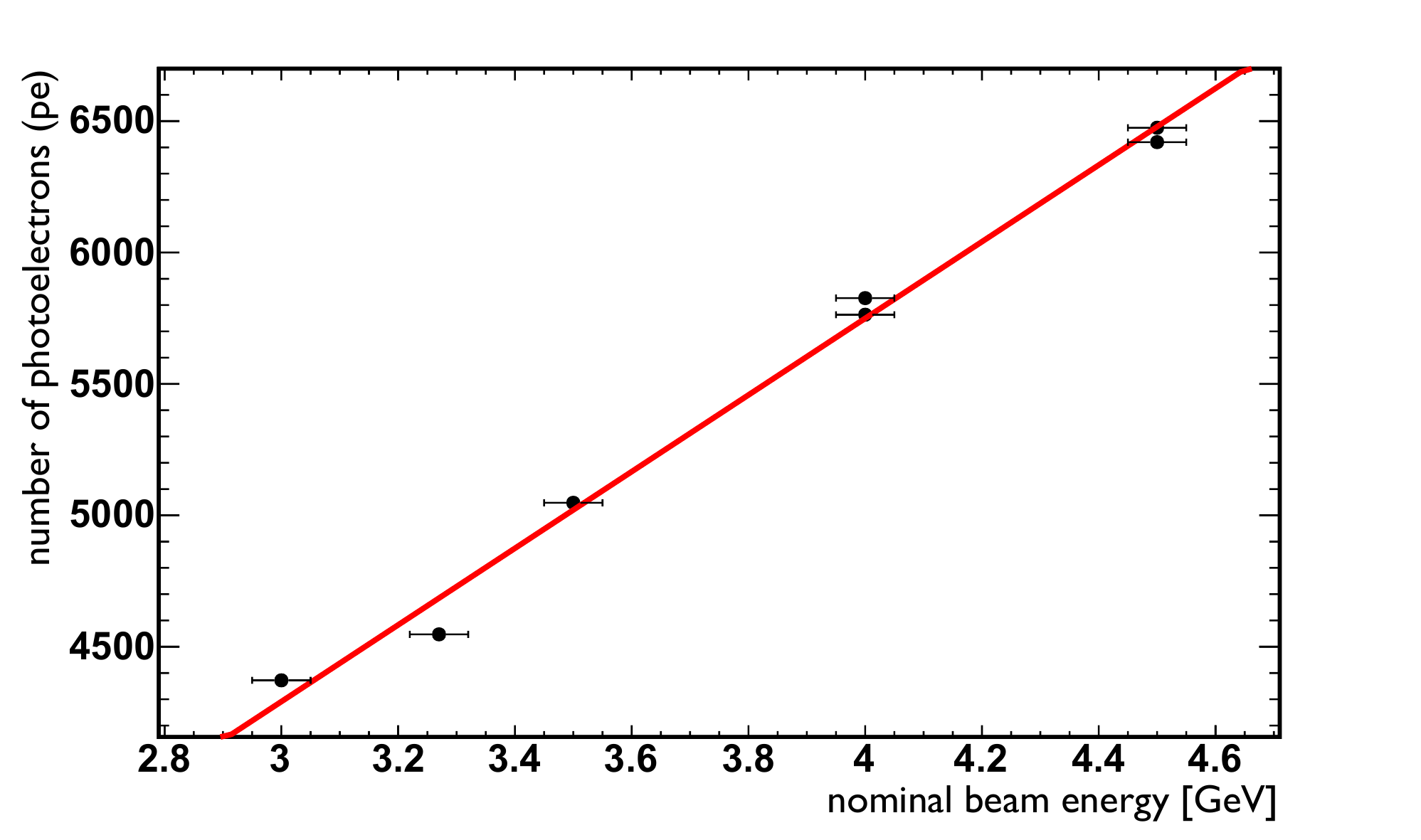}

   \caption{Measured pe in a 3\,$\times$\,3 cluster of white-wrapped crystals centered on SiPM~\#11 as a function of nominal beam energy. The uncertainties on the nominal beam energies are set to 50\,MeV.}
   \label{fg:ENERGY_LINEARITY}
\end{figure}

\section{Results}
\subsection{Energy Response and Light Yield}
\label{section:ENERGY_RESPONSE}

The energy response, light yield, and resolution of the detectors were studied for 3\,$\times$\,3 clusters of white- and black-wrapped crystals using a beam with energies ranging from 2 to 4.5\,GeV. SiPM~\#11 was the designated center of the white cluster and SiPM~\#24 was the center of the black array.

Offline analysis revealed that one of the quadrants in SiPM~\#24---the central SiPM in the black array---was not functioning and therefore was not being included in the final stage of amplification.
This amounted to an effective decrease in coverage of the central crystal in the black cluster to 75\,\% of what it would have been had the SiPM been operating correctly.
All pe values calculated for SiPM~\#24 were increased by a factor of 4/3 to correct for this effect.

\begin{figure}[t]
   \includegraphics[width=1.0\columnwidth]{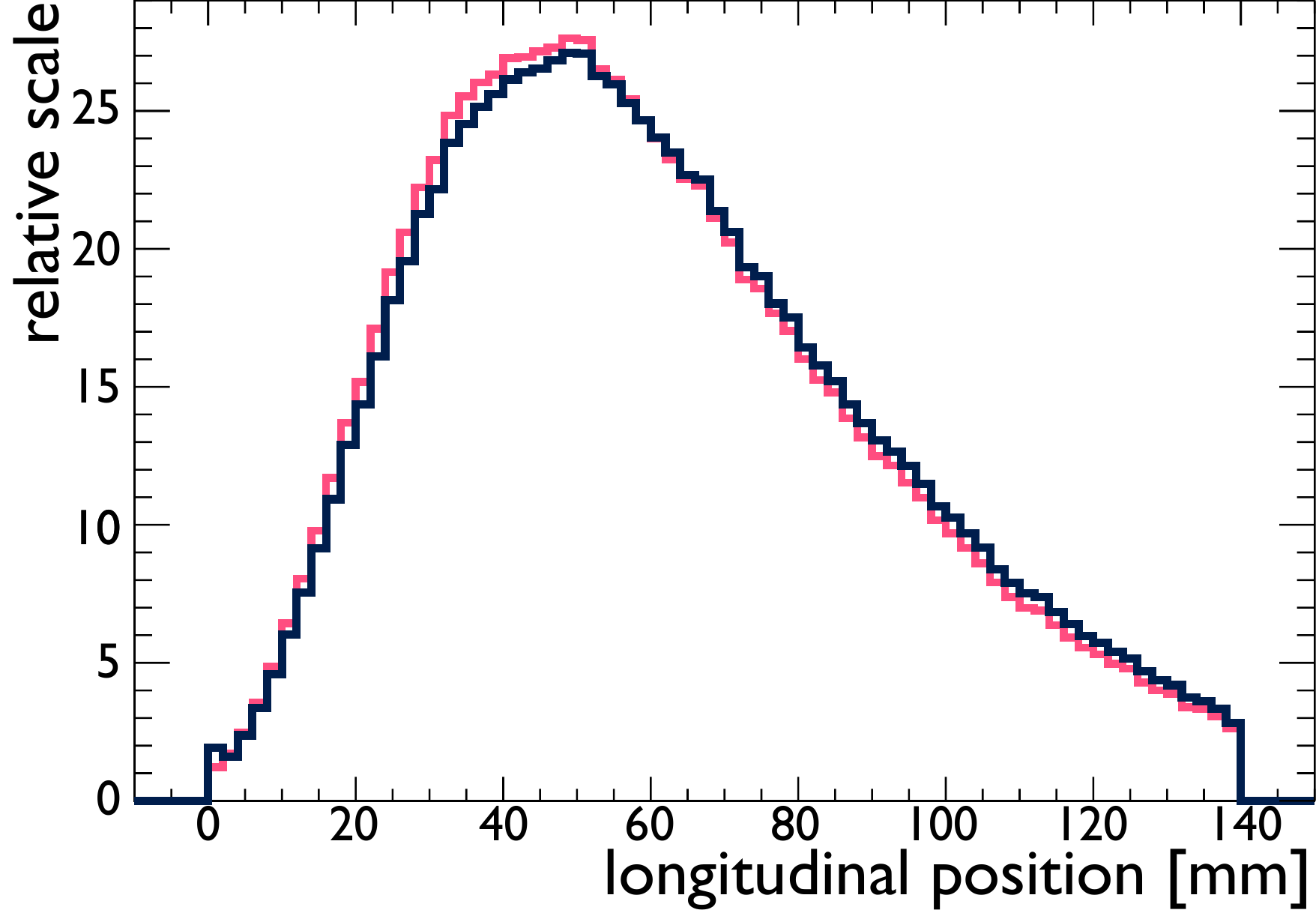}

   \caption{Simulated longitudinal profile of an electromagnetic shower deposited by a 3-GeV electron in a 140\,mm thick slab of \pb. Energy deposits in dark blue and generated \v{C}erenkov photons in light red.}
   \label{fg:energy-depostion-longitudinal}
\end{figure}

For every beam event, the pulses in each of the nine crystals within a cluster were converted into pe and summed together. As shown in Fig.~\ref{fg:SAMPLE_FIT}, the total pe distribution is not symmetric. A low-energy tail exists owing to incomplete longitudinal shower containment.  The effective mean values are extracted from Gaussian fits that include an asymmetric fit region as shown in the figure.

A linear relationship between pe extracted from the fit and the beam energy in the range%
\footnote{Operating the accelerator below 3\,GeV required very low magnet currents and reproducing energy setting suffered from  hysteresis effects. As a consequence, the energy uncertainty associated with these data points was too large for the points to contribute meaningfully to the linearity fit, and the points were not used.}
from 3\,--\,4.5\,GeV is shown in Fig.~\ref{fg:ENERGY_LINEARITY}.
We obtained a slope of $(1.45 \pm 0.05)$\,pe/MeV with an offset of $(-80 \pm 200)$\,pe for the white array and a slope of $(0.76 \pm 0.04)$\,pe/MeV with an offset of $(-150 \pm 160)$\,pe for the black array%
\footnote{One corner crystal from the black array---which contains less than 1\,\% of the total energy deposition---was not included in the sum because of a lack of available digitizer channels.}.

\subsection{Energy Resolution}

According to our \textsc{Geant4} simulations, an electron in the energy range 2\,--\,4.5\,GeV will deposit $(96.3 \pm 1.0)$\,\% of its energy in a 140\,mm-deep, infinitely-wide block of \pb.
The uncertainty was determined from a Gaussian model fit to the MC generated histogram of the deposited energies in an asymmetric region, identically to how the experimental pe histograms was handled in Fig.~\ref{fg:SAMPLE_FIT}. The method is motivated by a direct comparison to the literature at the end of the section.
 A longitudinal profile of the deposited energy and generated \v{C}erenkov photons is shown in Fig.~\ref{fg:energy-depostion-longitudinal}. They are subtly different. As the shower develops and propagates deeper into the crystal, secondary charged particles are emitted with lower mean energies. That means the kinetic energy of these particles approaches the \v{C}erenkov threshold, and a smaller fraction of \v{C}erenkov photons is emitted per unit of energy lost. Because the tail of the shower contains mainly lower-energy particles, the relative variance in the \v{C}erenkov photon yield is greater compared to the upstream shower contribution.

\begin{figure}[t]
   \includegraphics[width=1.0\columnwidth]{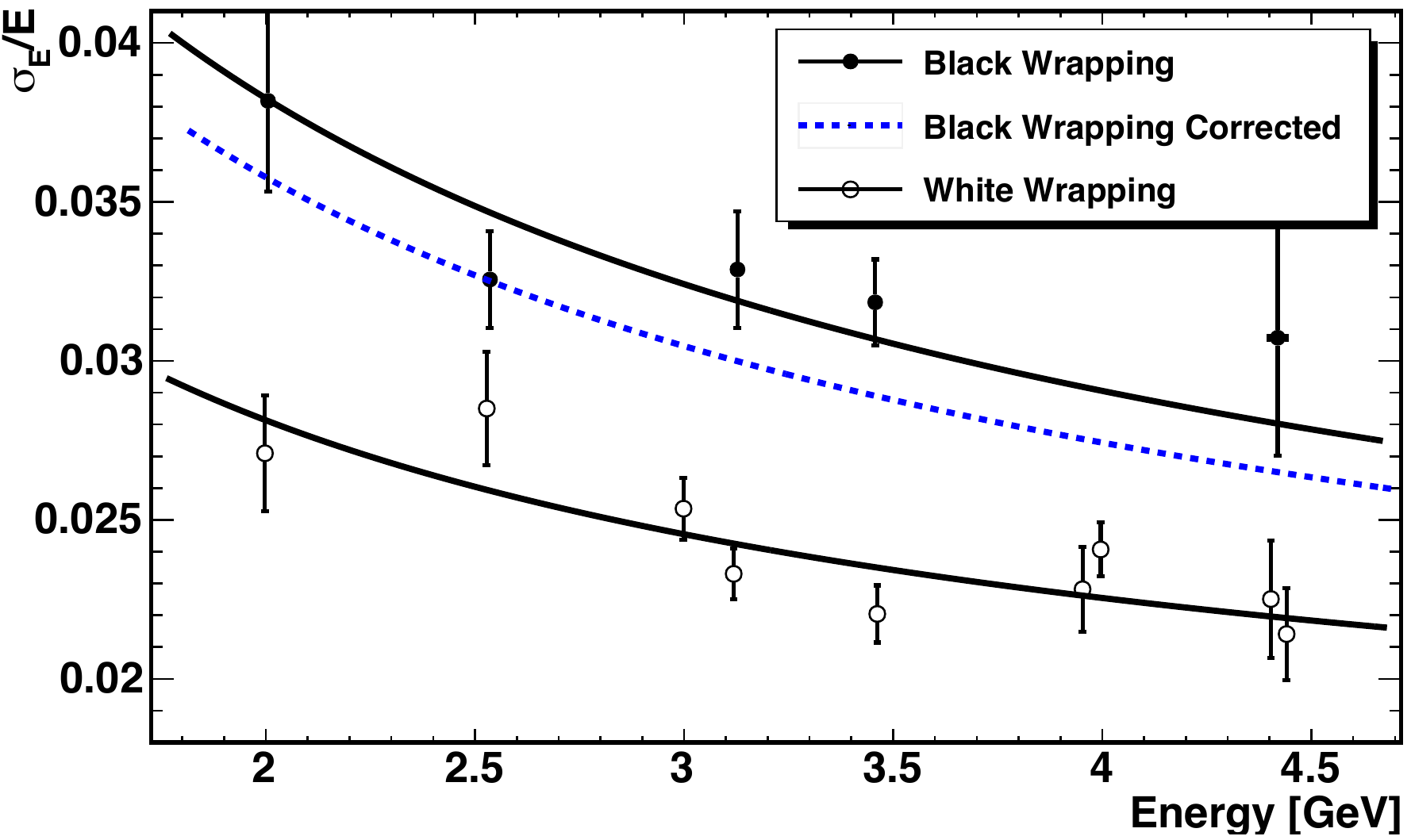}

   \caption{Energy resolutions of 3\,$\times$\,3 arrays of \pb~crystals with black and white wrappings as a function of energy.
    Fit functions are of the form $\sigma^2_E/E^2 = (1.5\,\%)^2 + a^2/(E/\mathrm{GeV})$.
    The blue dashed line is the result of correcting the black-wrapped curve for dead SiPM channels, as described in the text.}
   \label{fg:ENERGY_RESOLUTION}
\end{figure}

For the determination of energy resolution, one must properly account for fluctuations in the containment fraction for a finite-sized detector. The 1\,\% variance predicted from the simulation for {\em energy} containment increases to 1.3\,\% when one is considering generated photons, which is what is counted by the SiPM.  If we next consider the transverse containment of the 3\,$\times$\,3 cluster, we find that the fluctuations from the 2.7\,\% fraction of \v{C}erenkov photons generated outside of the array further increases the overall containment-based variance to 1.5\,\%.  Thus, the constant containment-based variance must be accounted for in the resolution function as a term behaving as: $\sigma_E/E = 1.5\,\%$.

Assuming all other contributions to the energy resolution scale with the energy in the same way as the number of pe, i.e., $1/\sqrt{E/\mathrm{GeV}}$, the complete energy resolution expression can be described by
\begin{equation}
  \frac{\sigma_E}{E} =  \sqrt{(1.5\,\%)^2 + \frac{a^2}{E/\mathrm{GeV}}} \quad.
\label{eq:eresolution}
\end{equation}
Fits to this function yield $a = (3.4 \pm 0.1)$\,\% with a reduced $\chi^2$ of 10.6/8 for the white array and $a = (5.0 \pm 0.3)$\,\% with a reduced $\chi^2$ of 3.5/4 for the black array (see Fig.~\ref{fg:ENERGY_RESOLUTION}).
The experimental data points in Fig.~\ref{fg:ENERGY_RESOLUTION} were extracted from the pe distributions (Fig.~\ref{fg:SAMPLE_FIT}) using the same procedure as for the mean values, i.e., the point is the width parameter of the Gaussian fit in an asymmetric region.

\begin{figure}[t]
   \includegraphics[width=1.0\columnwidth]{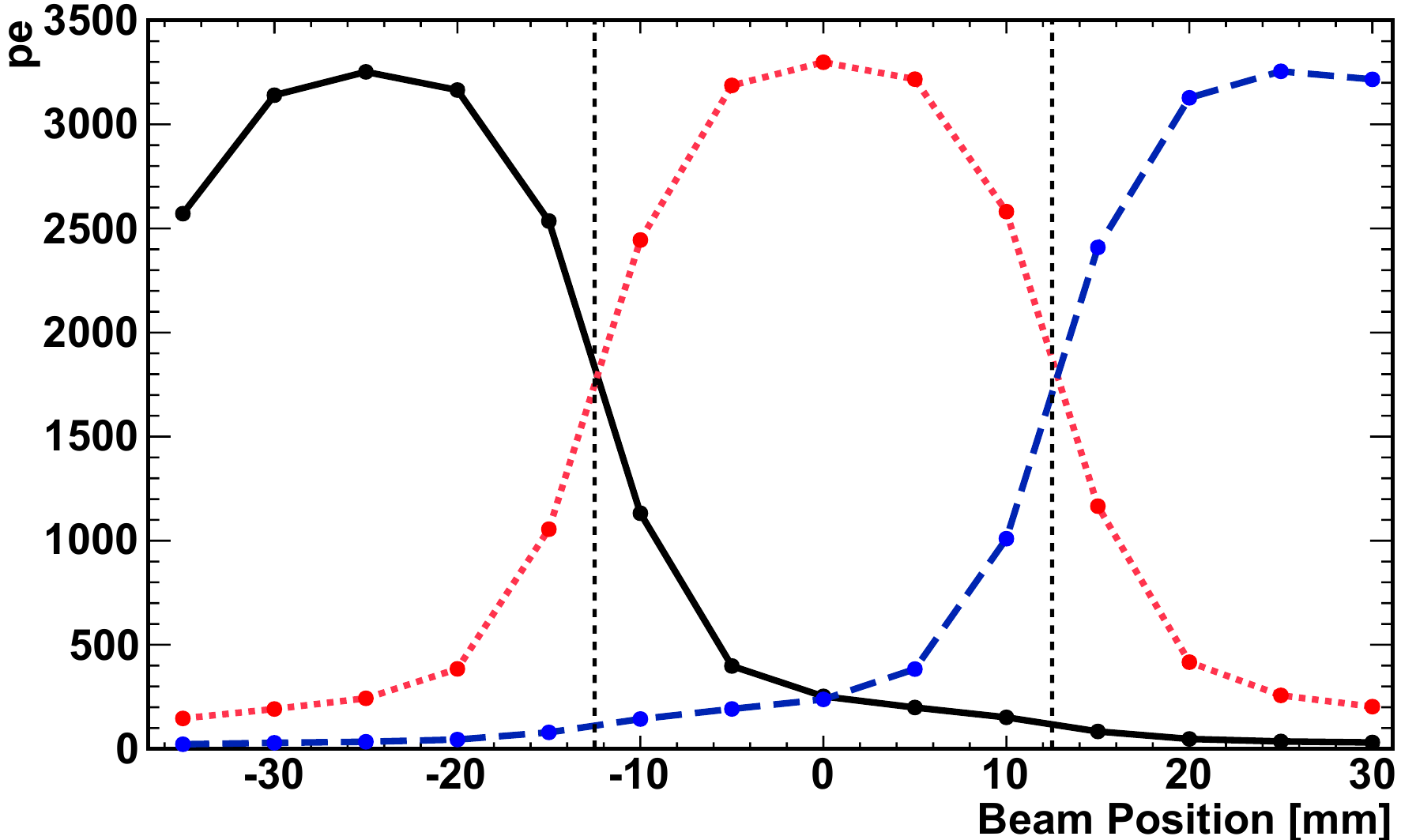}
   \caption{The number of pe detected in each of three adjacent white-wrapped crystals as a 3.1\,GeV electron beam is scanned across them.
     Dotted lines are drawn along the crystal boundaries.}
   \label{fg:POSITION_SCAN_PE}
\end{figure}

The obtained values of the $a$ term are dominated by the photo-statistics contributions predicted from the light-yield values. The remaining parts can be assigned to uncertainties in the calibration of various SiPMs together with position fluctuations of the beam, and inhomogeneous response of the SiPMs themselves. In principle the momentum uncertainty, $\Delta P/P$, of the  beam can also contribute to the constant term, which we held fixed in the fit.  SLAC asserted that $\Delta P/P < 1\,\%$.

The constant term value of 1.5\,\% was obtained from the MC simulation in the same way as the experimental data were processed: A Gaussian model was fit in an asymmetric region of the MC generated pe histogram. In general, fixing the constant term might pose a systematic bias. A fit to the experimental data with a free constant term was performed to investigate the possible systematics. If the constant term is allowed to float, the fit prefers the value of $(1.6 \pm 0.3)$\,\%. The difference in $\chi^2$ values is vanishing between the two fits and does not justify adding the extra free parameter. However, a non-vanishing constant term is essential for a successful fit. If the constant term is fixed to zero, the $\chi^2$ value increases from 11 to 19 which for 8 degrees of freedom corresponds to the $p$-value of 1.7\,\%.

\begin{figure}[t]
   \includegraphics[width=1.0\columnwidth]{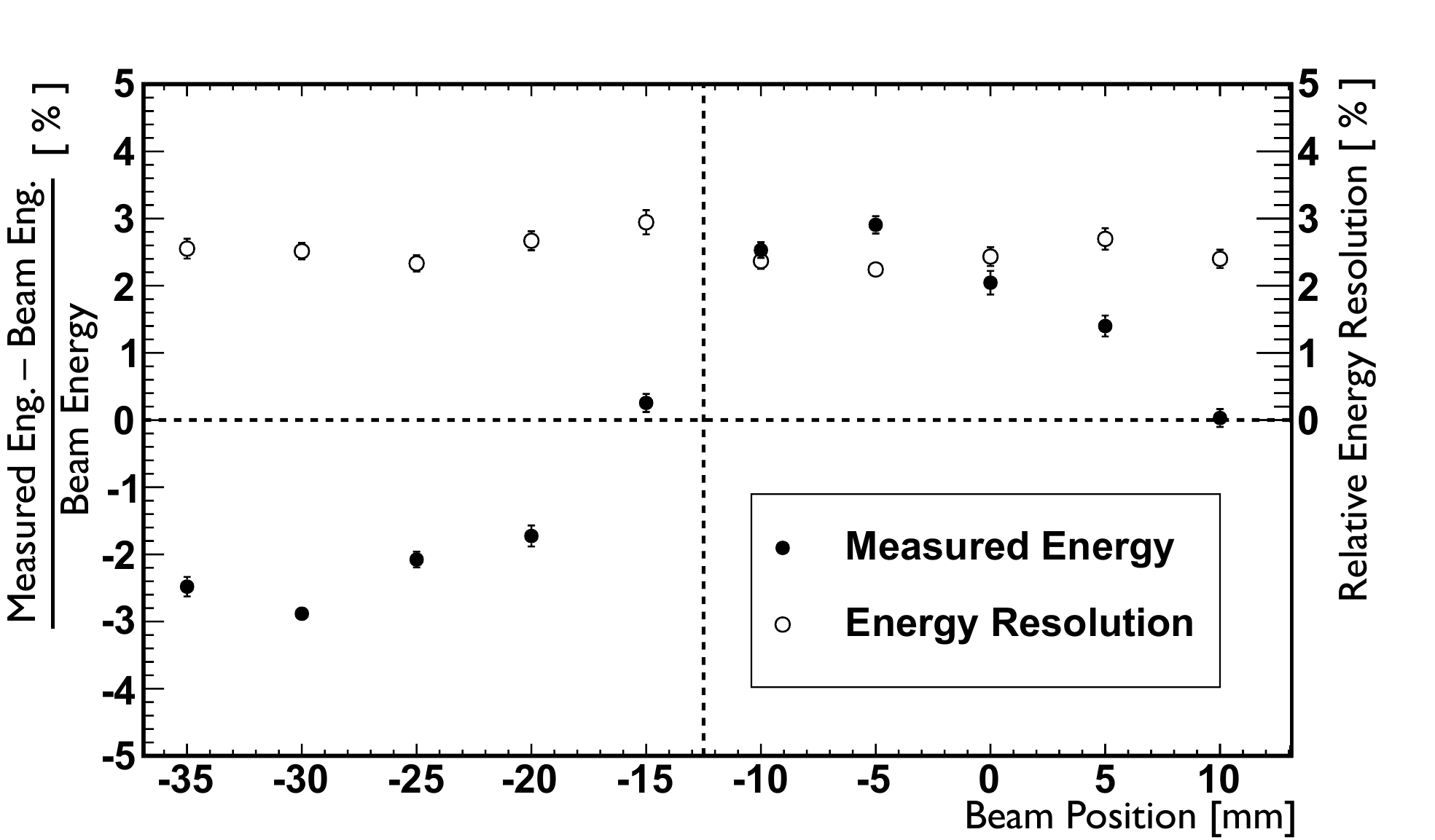}
   \caption{The measured energy and energy resolution in a 3\,$\times$\,4 array of white-wrapped crystals during a position scan with a 3.1-GeV beam.
   A vertical dotted line is drawn along a crystal boundary and a horizontal dotted line is drawn along the mean measured energy of 3.1\,GeV.}
   \label{fg:POSITION_SCAN_ENG}
\end{figure}

The energy resolution measurement for the black-wrapped array suffered because of the dead quadrant in SiPM~\#24; lower crystal coverage results in an increased variance in the photo-statistics term.
While we can correct the light yield by increasing the measured pe in SiPM~\#24 by 4/3, the increased weighting does not account for the increased photo-statistics contribution to the relative variance, which behaves as $1/N_\mathrm{pe}$, where $N_\mathrm{pe}$ is the mean number of photoelectrons detected by the array. 

\begin{table}[b]
  \centering
  \begin{tabular}{|p{0.37\columnwidth}||c|c|}\hline
     {} & Millipore & Tedlar \\ \hline
    Light yield \newline [pe/MeV] & $1.45 \pm 0.05$ & $0.76 \pm 0.04$ \\ \hline
    Energy resolution\newline [$\%/\sqrt{E/\mathrm{GeV}}$] & $3.4 \pm 0.1$ & $4.6 \pm 0.3$ \\
    \hline
  \end{tabular}
  \caption{Summary of results for 3\,$\times$\,3 array. Note the Tedlar-wrapped array energy resolution has been corrected following the procedure described in text.}
  \label{table:ENERGY_RESPONSE}
\end{table}

The correction process to account for the dead quadrant can be separated into two conceptual steps.  First, the obtained photo-statistics contribution is subtracted quadratically from the energy resolution.  Next, the expected photo-statistics contribution is added back in. Let $L_\mathrm{tot}$ be the total light yield of the 3$\times$3 array in pe/GeV and $f_c$ the fraction of $L_\mathrm{tot}$ collected in the central crystal with a fully working SiPM. The value for $f_c$ in the black-wrapped array is 0.85, and $L_\mathrm{tot}$ is 760\,pe/GeV.
The obtained photo-statistics contribution to the relative variance when 3/4 of the central SiPM is in operation is $[EL_\mathrm{tot}(\frac{3}{4}f_c + 1 - f_c)]^{-1}$.  The expected  contribution if all channels are operating is $(EL_\mathrm{tot})^{-1}$. Because the photo-statistics contribution to the relative variance is of the form $E^{-1}$, the described procedure corrects $a$ in the energy resolution expression (Eq.~\ref{eq:eresolution}).

Therefore, the result $a_\mathrm{corrected} = (4.6 \pm 0.3)\,\%$ is obtained. A summary of the results from this and the previous section are given in Table~\ref{table:ENERGY_RESPONSE}.
Despite the SiPMs' small area coverage, the energy resolutions obtained here are better than those measured in previous studies of \pb\ using photo-tubes~\cite{Anderson:1989uj, Appuhn:1993zu, Baunack:2011xva}.

\begin{figure}[t]
   \includegraphics[width=1.0\columnwidth]{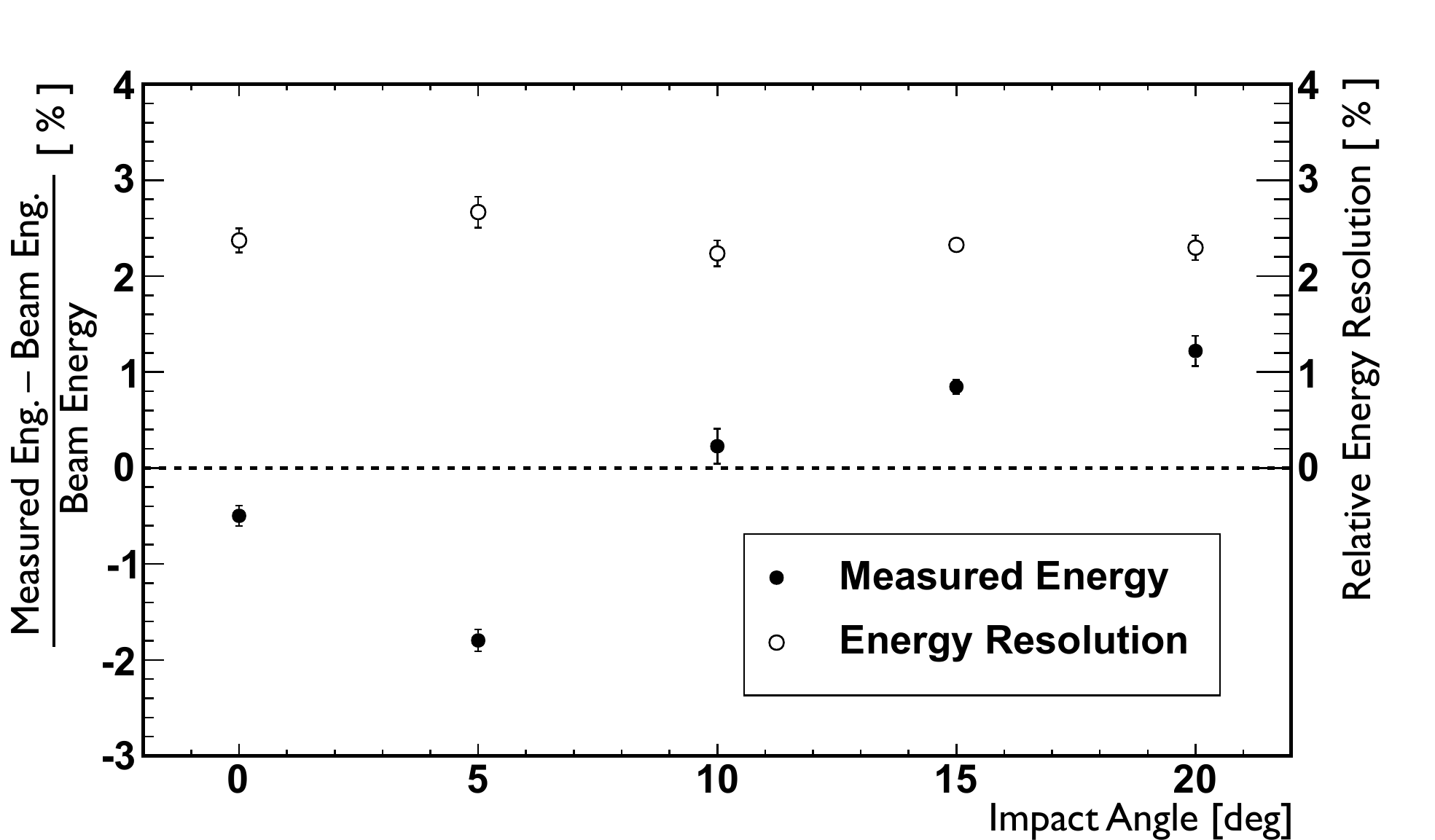}
   \caption{The measured energy and energy resolution in a 3\,$\times$\,4 array of white-wrapped crystals as a function of the impact angle for a 3.0-GeV electron.
     The impact angle was varied by physically rotating the calorimeter.
     A horizontal dotted line is drawn along the mean measured energy of 3.0 GeV.
     Note: The resolutions are slightly better than those described earlier because here the beam is  centered on SiPM~\#6 rather than \#11, and SiPM~\#6 had a higher light yield.
   }
   \label{fg:ANGLES}
\end{figure}

\subsection{Impact Position}

A 3.1-GeV electron beam was scanned horizontally across the centers of three adjacent crystals in steps of 5\,mm. The behavior of the number of detected pe's (Fig.~\ref{fg:POSITION_SCAN_PE}) and the total pe sum  across the scan (Fig.~\ref{fg:POSITION_SCAN_ENG}) demonstrate the efficacy of the laser calibration.  The maximum numbers of pe observed in each crystal differ by under 2\,\%, while the total sum over 12 crystals remains constant within 4\,\% over the 45\,mm scan. Additionally, the energy resolution remains acceptable when the beam is placed near the boundary of two crystals. The slight differences in collected in the number of collected photo-electrons as the beam is moved from one crystal to the next can be explained by small differences in light-yield between crystals.

A previous study has obtained a position resolution of $\sigma_{x,y} = (0.99 \pm 0.06)$\,mm for a 2\,$\times$\,2 array of \pb\ crystals with dimensions 21\,$\times$\,21\,$\times$\,175\,mm$^3$~\cite{Appuhn:1993zu}.
We expect our calorimeter to exhibit similar position resolution, but no independent reference detector was available  so we were unable to confirm this expectation.

Nevertheless, the electron impact position was calculated from shot-to-shot using a data-driven empirical function of the center of mass of the crystal with the largest energy deposit and its nearest neighbors, where mass refers to the number of pe collected from a crystal. The impact position obtained using this technique was distributed with a width of $(2.5 \pm 0.1)$\,mm, which can be taken as an upper bound on the beam size. This result is statistically compatible with the expected radial profile of the electron beam.

\begin{figure}[t]
   \includegraphics[width=1.0\columnwidth]{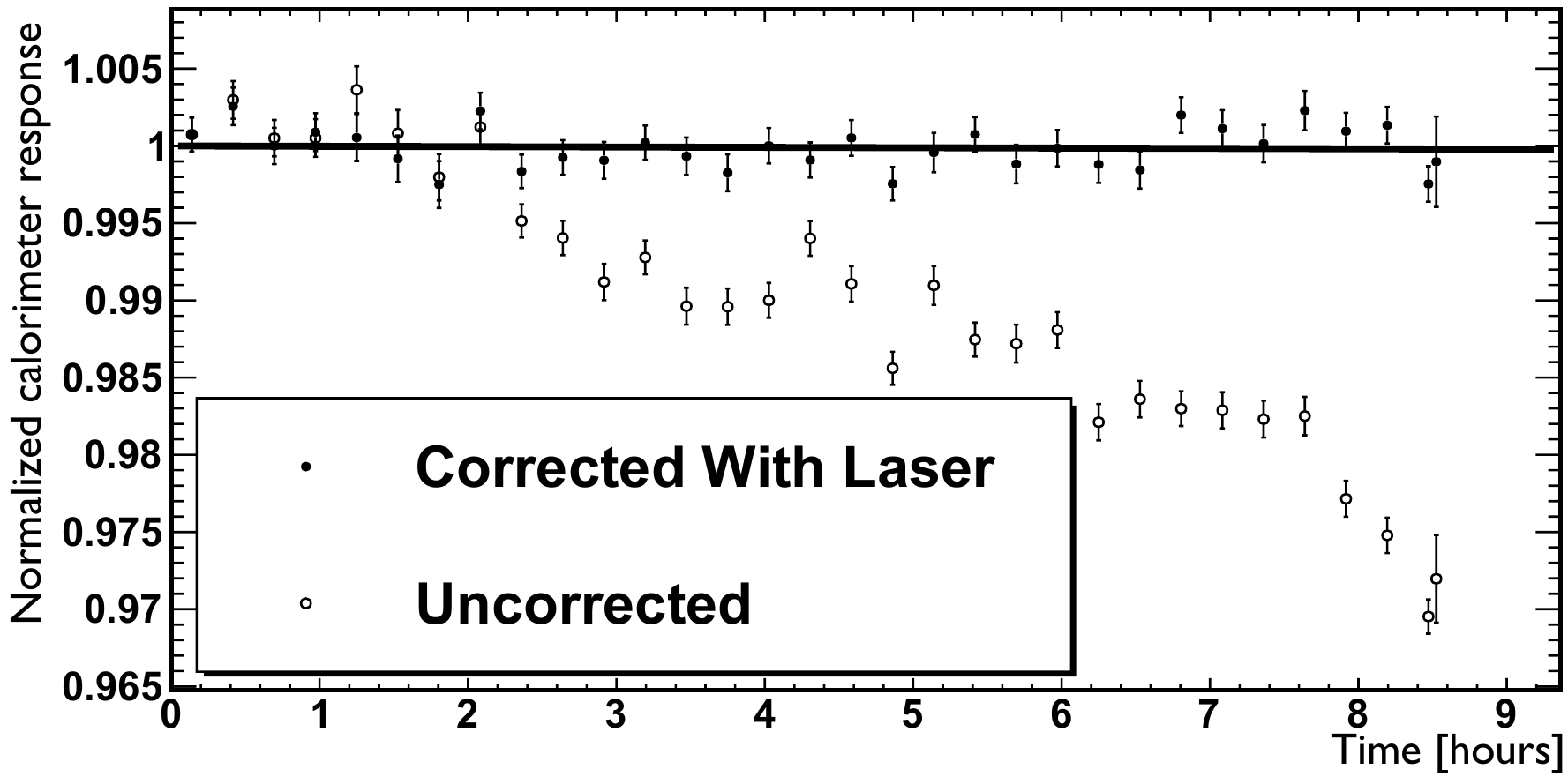}
    \caption{Measured energy in a 3\,$\times$\,4 white-wrapped array with a 3-GeV electron beam centered on SiPM~\#10, normalized to 1 at $t=0$, over a 9-hour period of constant running.
     The empty points are uncorrected for gain drift whereas the full points are corrected using the laser calibration system.
     The laser system was firing throughout this entire period, allowing for continual gain corrections.
   }
   \label{fg:STABILITY}
\end{figure}

\subsection{Impact Angle}

Measuring the response of the calorimeter for incident particles striking at non-normal angles was an important check because, in the \gm~experiment, decay positrons of interest curl inward from the storage ring and strike the calorimeter front face at energy-correlated angles in the range from 0 to  $\sim\!30$ degrees. This does not present a leading-order systematic problem, but we needed to record the calorimeter response to calibrate our simulation models.

We studied incident electrons with angles up to 20 degrees by physically rotating the calorimeter with respect to the beamline.
Angles of 0$^\circ$, 5$^\circ$, 10$^\circ$, 15$^\circ$, and 20$^\circ$ were tested with a 3-GeV beam centered on SiPM~\#6.
The results of these measurements are shown Fig.~\ref{fg:ANGLES}: the reconstructed energy is constant within 3\,\% and the energy resolution does not appear to be significantly affected by the electron angle.

\begin{figure}[t]
   \includegraphics[width=1.0\columnwidth]{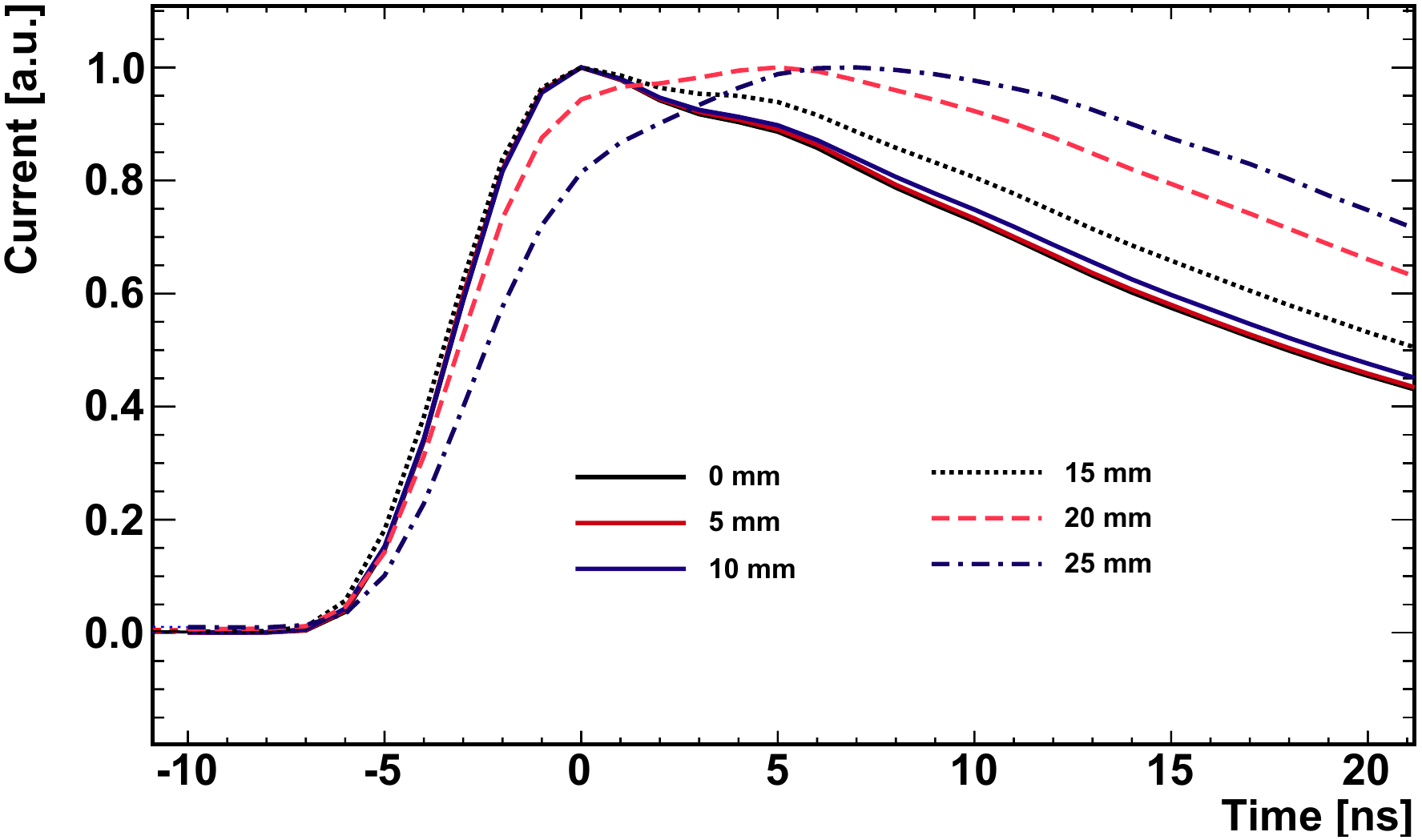}
   \includegraphics[width=1.0\columnwidth]{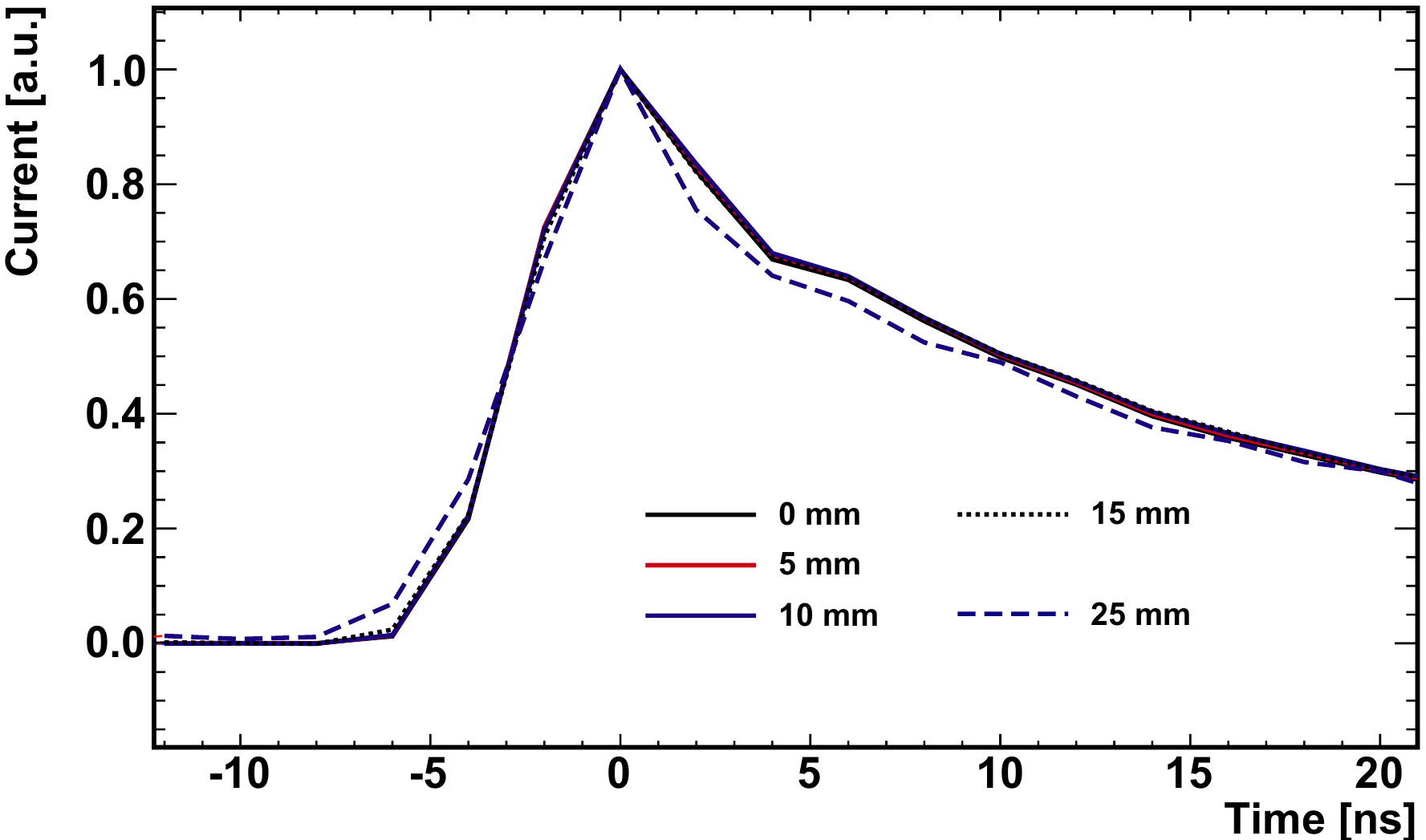}
   \caption{Pulse-shape vs. impact position difference from the crystal front face center. \emph{Top:}  White wrapping. For the (0, 5, 10\,mm) impact positions, the pulse shapes are practically identical, but for larger impact positions the pulse shape widens considerably.
   \emph{Bottom:} Black wrapping.  The pulse shape does not change significantly.}
   \label{fg:PULSESHAPE_VS_DISTANCE}
\end{figure}

\begin{figure*}[t]
    \includegraphics[width=\textwidth]{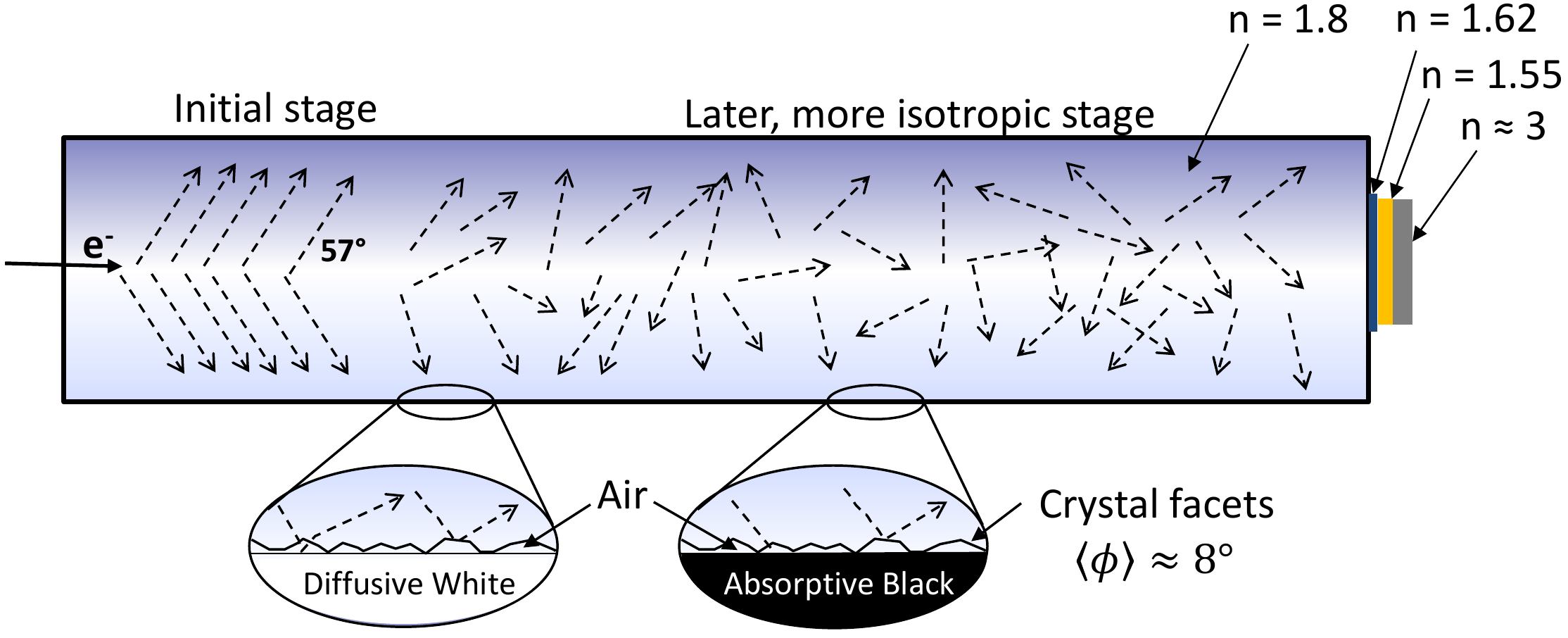}
\caption{Sketch of representative features of the light propagation in the crystal.  Near the entrance, the \v{C}erenkov cone produces photons at 57$^\circ$ with respect to the crystal axis which is aligned with the beam direction. As the shower develops, the average angle loses its original coherence, becoming nearly isotropic closer to the end of the crystal.  As the lower expansions illustrate, light bounces off of the faceted crystal surface owing to internal reflection for angles less than the critical angle $\theta_c = 90^\circ - 57^\circ = 33^\circ$. For larger angles, it passes into a thin layer of air and then hits the wrapping.  We assume the black wrapping to be totally absorptive, while the white wrapping is diffusive, re-emitting the light at a new angle, that might again be captured in the crystal.  The indexes of refraction of the crystal (1.8), optical gel (1.62), protective epoxy resin (1.55), and SiPM active surface ($\sim\!3$) are shown.  Only photons passing through this sequence can convert to photoelectrons.}
    \label{fg:cartoon}
\end{figure*}

\subsection{Long-term Gain Stability}

Long-term gain stability is a concern with SiPMs because of their sensitivity to changes in temperature.
Over the course of the run, the drifts in gain were tracked using the laser calibration system, which ran in parallel to normal data taking using a separate trigger at roughly 20\,Hz.
Fig.~\ref{fg:STABILITY} shows the reconstructed electron energy for a 3\,$\times$\,4 array of white-wrapped crystals using a 3-GeV beam over a 9-hour period binned in 15-min intervals.  The open circles represent the raw sum, which drifts downward by several percent during the study.  The solid circles are corrected using the information from laser events. First, an average response is determined on a shot by shot basis by using the sum of the 12 individual normalized crystal responses.  Each 15-minute period raw data histogram is then corrected for by the average laser response over this same period.  With the correction, $\delta E/E$ is stable at the level of $(2 \pm 9) \times 10^{-5}$ per hour.

\section{Pulse Shape vs. Impact Position}

\begin{figure}[t]
   \includegraphics[width=1.0\columnwidth]{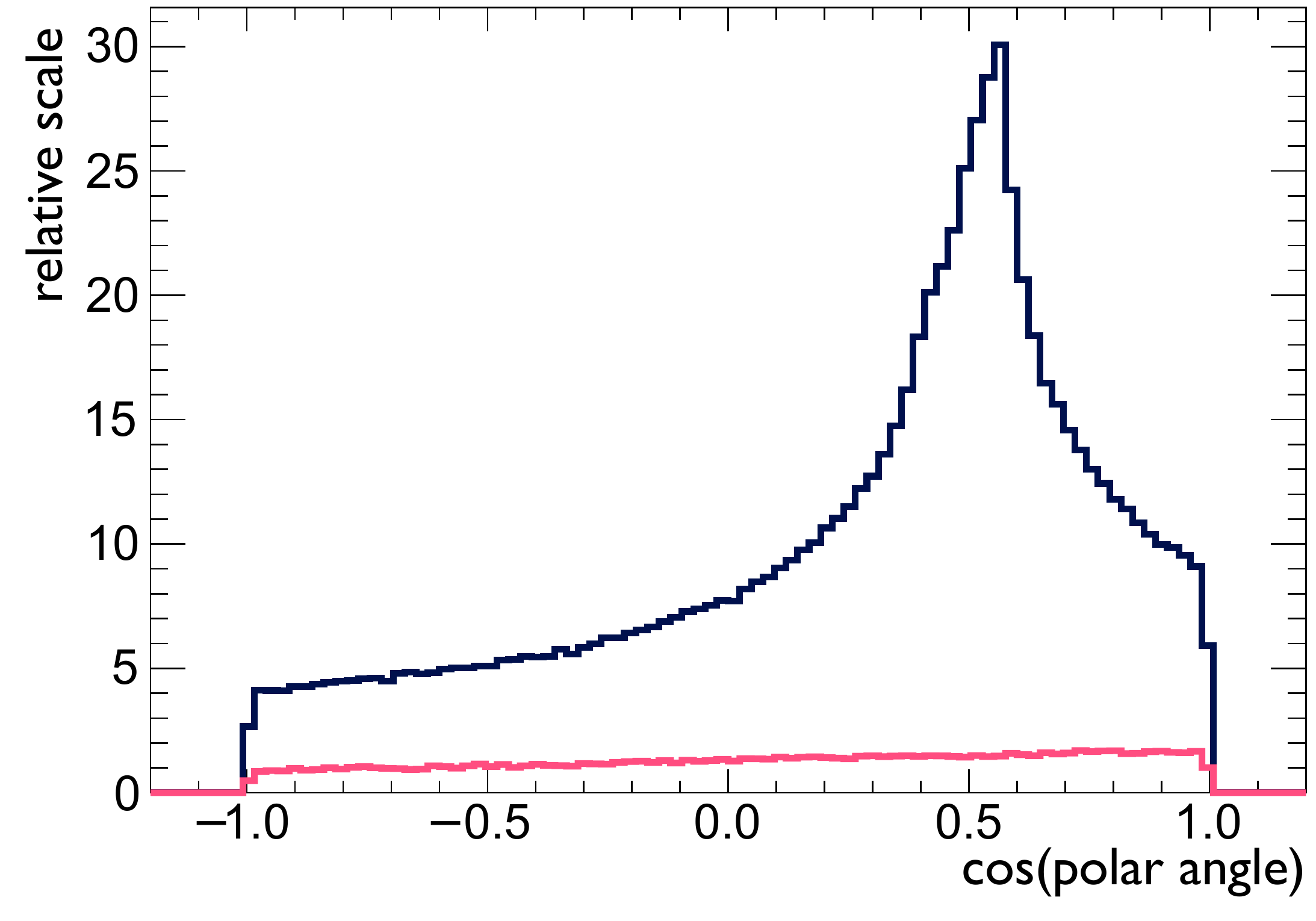}

   \caption{Angular distributions---with respect to the crystal longitudinal axis---of \v{C}erenkov photons produced by a 3-GeV electron in central (dark blue) and neighbor (light red) crystals.
     In the central crystal the distribution peaks at the \v{C}erenkov cone angle of 57$^\circ$ whereas in the neighbor crystals the distribution is nearly isotropic.
   }
   \label{fg:photon-angular-distribution}
\end{figure}

In this section we discuss the important finding that the pulse shape varies with impact position for the white-wrapped crystals, but not for the black-wrapped crystals, as shown in Fig.~\ref{fg:PULSESHAPE_VS_DISTANCE} and quantified in Table~\ref{table:FWHM}. This fact motivated a detailed Monte Carlo investigation to track the generation and propagation of \v{C}erenkov photons through accurate models of the crystals, respecting their faceted surfaces, and modeling the separate wrapping materials. Photons converted in the SiPM must have trajectories permitting them to exit the \pb\ material ($n = 1.8$), pass through an optical gel ($n = 1.62$), and a protective epoxy resin ($n = 1.55$), before striking the silicon surface.  Fig.~\ref{fg:cartoon} illustrates many of the key considerations that are included in the detailed light propagation simulation.  Fig.~\ref{fg:photon-angular-distribution} shows the net angular distribution of \v{C}erenkov photons produced in a central crystal vs. those produced in its nearest neighbor. The \v{C}erenkov cone is non-existent for the neighboring crystals since energy deposited there emerges from deeper in the shower where the direction of the deposited energy is no longer aligned with the direction of the incoming electron.

Let us first consider a black-wrapped crystal where the propagation of light is governed only by total internal reflection.  Photons that are accepted into the photo-detector have an angular range set by the index of refraction of \pb, with a sharp cut-off smeared by the Fresnel law, and by the angular distribution of the crystal surface facets.
The time to hit the crystal readout face and the detection probability once arrived naturally depend on where photons are produced and their angular distribution.  At the start of a shower---far from the readout surface---photons are largely generated in the \v{C}erenkov cone of $\sim\!57^{\circ}$.  Deeper into the shower---and closer to the readout surface---the photon angles develop an isotropic component. Because of the hard angular limit for captured photons, the acceptance vs. depth-of-generation is not uniform.

The number of photons emerging from the coherent part of the \v{C}erenkov cone in the initial phase of a shower is relatively constant vs. impact position, since this stage of the shower is narrow transversely.   As the shower evolves deeper and widens, the relative number of the more ``isotropically'' produced photons will depend on impact position, following the general sharing of energy between neighboring crystals. This asymmetry provides a mechanism for a possible pulse-shape evolution if the two sources of photons are treated differently or collected differently in time.  Because the black-wrapped crystal is dictated only by total internal reflection, both the central crystal and the neighbor pulse shapes are close to identical and neither depends perceptibly on impact position.  The mean arrival time differs slightly because the shower propagates faster than the light does, and thus the mean arrival time is about 200\,ps earlier for a neighbor crystal compared to a central one, in general.

\begin{table}[t]
  \centering
  \begin{tabular}{|p{0.30\columnwidth}||c|c|}
    \hline
    Distance from & \multicolumn{2}{c|}{Relative FWHM}\\ \cline{2-3}
    Beam [mm] & Millipore & Tedlar \\ \hline
    0 & $1$ & $1$ \\
    5 & $1.01 \pm .01$ & $1.01 \pm .02$ \\
    10 & $1.05 \pm .02$ & $1.02 \pm .02$ \\
    15 & $1.17 \pm .01$ & $1.02 \pm .03$ \\
    20 & $1.40 \pm .01$ &  \\
    25 & $1.52 \pm .02$ & $0.96 \pm .05$ \\
    \hline
  \end{tabular}
  \caption{Pulse full width at half maximum evolution during position scan. The zero ``Distance from Beam'' refers to the beam hitting the center of a crystal. From that point, the scan is performed horizontally.}
  \label{table:FWHM}
\end{table}

White-wrapped crystals, in contrast, introduce an additional light propagation path, which allows photons generated with high polar angles to successfully reach a SiPM.  When a photon escapes the surface, it can be redirected into and recaptured by the crystal at a new angle. These photons have, on average, longer flight times prior to arriving at the photo-detector and thus both widen and increase the pulse compared to a black-wrapped crystal.  The width of the pulse shape corresponds directly to the width of the angular distribution, smeared by Lambertian reflection from the wrapping. As the impact position is displaced from the center, the average captured angular distributions change significantly and thus evolve the pulse width.

Our measurements show that the central crystal of a 3\,$\times$\,3 white-wrapped cluster contains  $(71 \pm 2)$\,\% of the detected photons, while for a black-wrapped cluster it contains $(85 \pm 2)$\,\% of the detected photons.  The efficiency of conversion of the upstream photons in the initial cone vs. those produced later and more isotropically is clearly wrapping dependent; for example, the neighbor crystal pulse is exclusively based on the ``more isotropic'' photons as shown in Fig.~\ref{fg:photon-angular-distribution}. To fully account for the intensity differences and the pulse width evolutions, our Monte-Carlo input parameters---average facet angle, crystal absorption length, and reflectivity of wrapping---were tuned using reasonable values close to expectations.  However, the solution is not unique and we plan to study these dependencies further.

\section{Conclusions}

We report on a study of an array of 28 \pb\ crystals read out using 16-channel, large-area SiPMs from Hamamatsu. This electromagnetic calorimeter represents a half-sized prototype for one of the 24 calorimeter units that are required for the new muon \gm\ experiment at Fermilab.  Unique features of this study include the exclusive use of fast waveform digitizers to record the pulses; the comparison of different crystal wrappings to investigate light yield, pulse shape, and energy resolution; and a laser-based calibration system with a high degree of pulse-to-pulse intensity stability.

Principle findings of this study are:
\begin{itemize}
\item The energy resolution, light yield, and linearity characteristics of a \pb\ calorimeter coupled with SiPM readout is found to either exceed or meet performance of previous PMT-coupled arrays.
\item The absolute energy scale in units of photo-electron per pulse-integral can be obtained using only the laser system, independent of beam, and the calibration system can monitor the gain to a relative precision of better than $10^{-4}$ per hour.
\item White-wrapped crystals exhibited an energy resolution $\sigma/E$ of  $(3.4\pm0.1)\,\%/\sqrt{E/\mathrm{GeV}}$, with nearly twice the light yield compared to black-wrapped crystals, that had a resolution of $(4.6\pm0.3)\,\%/\sqrt{E/\mathrm{GeV}}$. 
\item The crystal wrapping affects more than just the light yield; it affects the pulse-shape as a function of impact position, in particular with a white diffusive wrapping.
\end{itemize}

\section*{Acknowledgments}

We thank Adam Para for the initial suggestion to test \pb~as a candidate material and for excellent advice on evaluating and choosing SiPM products.  Tianchi Zhao negotiated the original contract with SICCAS and helped evaluate early samples.  Carsten Hast, Keith Jobe and Zenon Szalata hosted this effort at the SLAC ESTB facility, which is supported under Department of Energy (DOE) contract DE-AC02-76SF00515.
This research was supported by the National Science Foundation (NSF) MRI program (PHY-1337542), by the DOE Offices of Nuclear (DE-FG02-97ER41020) and High-Energy Physics  (DE-SC0008037), by the NSF Physics Division (PHY-1205792, PHY-1307328, PHY-1307196, DGE-1144153), by the Istituto Nazionale di Fisica Nucleare (Italy), and by the National Natural Science Foundation of China (11375115) and the Shanghai Pujiang Program (13PJ1404200).

\bibliography{pbf2_slac}

\end{document}